\documentclass[11pt]{article}
\usepackage[utf8]{inputenc}
\usepackage{amsmath, amssymb, amsfonts, xcolor, bbm, setspace, caption, hyperref, subfiles, chngcntr, natbib, enumitem, subfig, array, graphicx, amsthm, multirow, multicol, placeins, booktabs}

\allowdisplaybreaks

\theoremstyle{plain}

\newtheorem{proposition}{Proposition}
\newtheorem{assumption}{Assumption}

\theoremstyle{definition}

\newtheorem{remark}{Remark}

\theoremstyle{remark}

\renewcommand{\P}{\mathbb{P}} % probability
\newcommand{\R}{\mathbb{R}} % R 
\newcommand{\E}{\mathbb{E}} % expectation
\renewcommand{\text}[1]{\textnormal{#1}} % normal text

\newcommand{\supp}{\textnormal{supp}} % argmax

\begin{document}

\begin{titlepage}
\centering
\onehalfspacing

\title{Partial Identification of Individual-Level Parameters Using Aggregate Data in a Nonparametric Model\vspace*{10pt}}

\author{Sarah Moon\thanks{Department of Economics, Massachusetts Institute of Technology, Cambridge, MA 02139, USA. Email: \href{mailto:sarahmn@mit.edu}{sarahmn@mit.edu}} }

\maketitle

\abstract{
\noindent \selectfont  \noindent
I develop a methodology to partially identify linear combinations of conditional mean outcomes when the researcher only has access to aggregate data. Unlike the existing literature, I only allow for marginal, not joint, distributions of covariates in my model of aggregate data. Bounds are obtained by solving an optimization program and can easily accommodate additional polyhedral shape restrictions. I provide a procedure to construct confidence intervals on the identified set and demonstrate performance of my method in a simulation study. In an empirical illustration of the method using Rhode Island standardized exam data, I find that conditional pass rates vary across student subgroups and across counties.
\\[5pt]
}

\end{titlepage}

\onehalfspacing

\maketitle

\newpage

\setcounter{page}{2}

\section{Introduction}
\label{sec1}

It has been long known that the relationship between variables at the individual level can be different from the relationship between those same variables aggregated over individuals. The ecological inference literature, which presents potential solutions to this problem, dates back to the seminal work of \cite{robinson1950ecological}, \cite{duncan1953ecological}, and \cite{theil1954linear}. Much of the literature is concerned with point identification of individual-level parameters, which requires strong assumptions that may be implausible in applied settings (see, e.g., \cite{cho1998iff, cho2004limits, freedman1998solution} and \cite{kousser2001ecological}). Throughout this paper I use the phrase ``individual-level parameters'' to refer to parameters that rely on the joint distribution of the individual-level variables.

Absent such strong assumptions, I can still partially identify individual-level parameters based on aggregate data. The resulting identified set precisely quantifies the extent to which individual-level results are sensitive to assumptions.
In this paper I build upon the methods of  \cite{cross2002regressions} and \cite{cho2008cross} to derive partial identification of individual-level parameters when only aggregate data is available. I consider a model of aggregate data where marginal distributions of covariates are observed together with a marginal average outcome across different groups. The individual-level parameters of interest are linear combinations of conditional mean outcomes $\E[Y_i|X_{1i},\dots, X_{Li}]$, which encompass parameters like average predictive effects. 
I construct bounds by solving an optimization problem that considers all joint distributions of individual-level variables that are consistent with the observed marginal information. The optimization problem formulation allows for easy accommodation of additional model restrictions.

I provide a consistent plug-in estimator for the bounds and a valid, albeit conservative, inference procedure that works with aggregate data. I conduct a simulation study to demonstrate the performance of the estimation and inference procedures and what features of the aggregate data drive the width of the bounds. I find that when the aggregate data provides stronger restrictions on the joint distribution of covariates, bounds are more informative.

To examine the informativeness of these bounds in practice, I apply this methodology to a Rhode Island standardized exam dataset. 
I find that bounds are very wide on conditional pass rate gaps of interest. Imposing monotonicity shape restrictions narrows the bounds. With restrictions implied by additional subgroup pass rate data, bounds are more informative, especially under monotonicity and subgroup pass rate restrictions. A homogeneity restriction on the underlying individual-level model results in empty estimated identified sets. 

Relevant to this paper is the literature on partial identification when combining multiple data sets, also known as ecological inference, which deals with similar issues of needing to infer joint information that is unobserved \citep{cross2002regressions, molinari2006generalization, ridder2007econometrics, fan2014identifying, fan2016estimation, buchinsky2022estimation, haultfoeuille2024partial, elzayn2025monotone, mccartan2025identification}. Most relevant is \cite{cho2008cross}, who use the methods of \cite{cross2002regressions} to derive bounds on conditional mean outcomes when the observed data is the marginal distribution of $Y_i$ over groups together with the joint distribution of covariates $X_i = (X_{1i},\dots, X_{Li})$ for many groups. In contrast, in my model of aggregate data the researcher observes the marginal, not the joint, distribution of the covariates $X_{1i},\dots, X_{Li}$ for groups.
My model better fits the aggregate data seen in practice. For example, aggregate data can provide ethnicity distributions, gender distributions, education distributions, and income distributions but often does not provide the joint distribution of all of these variables due to privacy concerns. I argue that the existing \cite{cross2002regressions} and \cite{cho2008cross} methods do not immediately apply for the kind of aggregate data I consider. Furthermore, I demonstrate that approaches that are valid with aggregate data and directly use \cite{cross2002regressions} bounds produce bounds that are not sharp.

The rest of the paper proceeds as follows. Section \ref{sec2} presents identified sets on the parameters of interest.
Section \ref{sec3} presents a discussion of consistent estimation and inference procedures. In Section \ref{sec4} I perform three simulation studies to evaluate the performance of the inference procedure and explore what features of the data drive bound width.
Section \ref{sec5} presents an empirical illustration of the methodology.
Section \ref{sec6} concludes.

\section{Identification}
\label{sec2}

In this section I construct identified sets for linear combinations of conditional mean outcomes using aggregate data.
I model aggregate data as marginal distributions of variables for many different groups. Suppose there exists a sequence of latent random variables $(Y_i, X_{1i}, \dots, X_{Li}, G_i), i = 1, \dots, n$, where $Y_i$ with support $[y_\ell, y_u]$ denotes individual $i$'s outcome, $X_i \equiv (X_{1i}, \dots, X_{Li})'$ with support $\{x_k\}_{k=1}^K$ denotes individual $i$'s covariates, and $G_i$ with support $\{1,\dots, G\}$ denotes individual $i$'s group. Aggregate data identifies the average outcome within each group $\E[Y_i|G_i = g]$, the marginal distributions of each of the $L$ covariates within each group $\P[X_{\ell i} = x_{k,\ell}|G_i = g]$, and the relative size of each group $\P[G_i = g]$.
In practice aggregate data consists of the sample equivalents of these values. I deal with sampling uncertainty in Section \ref{sec3}.  

I thus impose the following assumptions on the random variables:
\begin{assumption} \label{ass:1}
\begin{enumerate}[label=(\roman*)]
    \item $Y_i$ is a random variable with bounded support $[y_\ell, y_u]$.\footnote{The assumption of bounded support of $Y_i$ is for tractability of computing the identified set. Similar to \cite{cross2002regressions}, the main results in this section will still hold with unbounded support.}
    \item $G_i$ is a discrete random variable with finite support $\{1,\dots,G\}$.
    \item $X_i$ is an $L$-dimensional discrete random vector with finite support $\{x_k\}_{k=1}^K$ for $K \geq 2$.\footnote{The case of continuous covariates is beyond the scope of this paper. The assumption $K \geq 2$ ensures the problem is not trivial.}
    \item The joint distribution of the random variables $(Y_i, G_i, X_i)$ is latent. Instead, the researcher observes $\E[Y_i|G_i = g], \P[X_{\ell i} = x_{k,\ell}|G_i = g],$ and $\P[G_i = g]$ for every $\ell = 1, \dots, L, k = 1, \dots, K, $ and $g = 1, \dots, G$. Furthermore, $n$ is observed.
\end{enumerate}
\end{assumption}

As an example, consider a dataset of standardized exam results and demographics. $G_i$ denotes student $i$'s school district, $Y_i$ is an indicator for whether student $i$ passed the exam or not, and $X_i$ are student $i$'s demographics, like race and socioeconomic status. The researcher observes (sample estimates of) the pass rate for every school district $\E[Y_i|G_i=g]$, marginal demographics of every school district $\P[X_{\ell i} = x_{k,\ell}|G_i = g]$, and the number of students enrolled in each school district, with which one can obtain $\P[G_i = g]$. Alternatively, $Y_i$ could be student $i$'s exam score, and the researcher observes (sample estimates of) the average exam score for every school district $\E[Y_i|G_i=g]$.

The goal is to construct bounds on linear combinations of $\E[Y_i|X_i = x_k]$ for given weights $\{\lambda_k\}_{k=1}^K$, $\sum_{k=1}^K \lambda_k \E[Y_i|X_i = x_k]$. For example, if the researcher is interested in the average predictive effect of changing $X_i$ from $x_{k_1}$ to $x_{k_2}$ on $Y_i$, the researcher can choose $\lambda_{k_2} = 1, \lambda_{k_1} = -1,$ and $\lambda_k = 0$ for all other $k$. 
Note that this expectation is taken over the population of individuals $i$; the number of groups $G$ is fixed.
I will construct identified sets using only what is observed in aggregate data.

Construction of the identified set relies on bounding joint distributions of variables using marginal distributions. The object of interest, linear combinations of $\E[Y_i|X_i = x_k]$, depends both on the joint distribution of covariates $X_i$ and on the joint distribution of outcome $Y_i$ with covariates $X_i$. Neither of these distributions are observed in aggregate data. Instead, I construct bounds on these joint distributions similar to \cite{cross2002regressions}, and discuss how they relate to the \cite{cross2002regressions} bounds in Remark \ref{rem:cm}. I first use the marginal distributions of the $L$ covariates $\P[X_{\ell i} = x_{k,\ell}|G_i = g]$ to obtain bounds on the joint distribution of the covariates $\P[X_i = x_k|G_i = g]$. I then use the average of the outcome $\E[Y_i|G_i=g]$ together with bounds on the joint distribution of covariates to obtain bounds on the conditional mean group outcomes $\E[Y_i|X_i = x_k, G_i = g]$. Finally, I aggregate over groups to obtain bounds on (linear combinations of) the conditional mean outcomes $\E[Y_i|X_i = x_k]$.

I can bound joint distributions using marginal distributions by applying the law of iterated expectations and the law of total probability. The following three equations must hold:
\begin{align} 
    \P[X_{\ell i} = x_{k,\ell}|G_i = g] &= \sum_{j=1}^K \mathbbm{1}\{x_{j,\ell} = x_{k,\ell}\} \P[X_{i} = x_j|G_i = g], \label{eq:lie_x_g_unknown} \\
    \E[Y_i|G_i = g] &= \sum_{k=1}^K \E[Y_{i}| X_{i} = x_k, G_i = g] \P[X_{i} = x_k|G_i = g], \label{eq:lie_y_g_unknown} \\
    \E[Y_i|X_i = x_k] &= \sum_{g=1}^G \P[G_i = g|X_i = x_k] \E[Y_{i}|X_i = x_k,G_i = g]. \label{eq:lie_y_x_unknown}
\end{align}
From Bayes' rule and the law of total probability I also know
\begin{align}
    \P[X_i = x_k] &= \sum_{g=1}^G \P[X_i = x_k|G_i = g]\P[G_i = g], \label{eq:lie_x} \\
    \P[G_i = g|X_i = x_k] &= \frac{\P[G_i = g]\P[X_i = x_k|G_i = g]}{\P[X_i = x_k]}. \label{eq:bayes_rule}
\end{align}
I can use equations \eqref{eq:lie_x} and \eqref{eq:bayes_rule} to rewrite \eqref{eq:lie_y_x_unknown} as 
\begin{align*}
    \E[Y_i|X_i = x_k] &= \sum_{g=1}^G \frac{\P[G_i = g]\P[X_i = x_k|G_i = g]}{\sum_{g=1}^G \P[G_i = g]\P[X_i = x_k|G_i = g]} \E[Y_i|X_i = x_k, G_i = g],
\end{align*}
which can be rearranged as
\begin{align}
    \E[Y_i|X_i = x_k] \sum_{g=1}^G \P[G_i = g]\P[X_i = x_k|G_i = g] &= \sum_{g=1}^G \P[G_i = g]\P[X_i = x_k|G_i = g] \E[Y_i|X_i = x_k, G_i = g]. \label{eq:lie_plug_in_unknown}
\end{align}
This rearrangement is equivalent when $\P[X_i = x_k] = \sum_{g=1}^G \P[G_i = g]\P[X_i = x_k|G_i = g] > 0$. If $\P[X_i = x_k] = 0$ then $\E[Y_i|X_i = x_k]$ is undefined. If $\E[Y_i|X_i = x_k]$ were to exist then the only information available is that it lies in $[y_\ell, y_u]$ by Assumption \ref{ass:1}.i. Thus I will work with equation \eqref{eq:lie_plug_in_unknown}, which is still well-defined and valid when $\P[X_i = x_k] = 0$ because both sides are zero when $\E[Y_i|X_i = x_k]$ is finite.

To make clear what is known and what is unknown under Assumption \ref{ass:1}, I will replace all unknown objects with variables in equations \eqref{eq:lie_x_g_unknown}, \eqref{eq:lie_y_g_unknown}, and \eqref{eq:lie_plug_in_unknown}. Let $p_{kg}$ denote $\P[X_i = x_k|G_i = g]$, $c_{kg}$ denote $\E[Y_i|X_i = x_k,G_i = g]$, and $d_{k}$ denote $\E[Y_i|X_i = x_k]$. Then I have the following three equations:
\begin{align} 
    \P[X_{\ell i} = x_{k,\ell}|G_i = g] &= \sum_{j=1}^K \mathbbm{1}\{x_{j,\ell} = x_{k,\ell}\} p_{jg}, \label{eq:lie_x_g} \\
    \E[Y_i|G_i = g] &= \sum_{k=1}^K c_{kg} p_{kg}, \label{eq:lie_y_g} \\
    d_k \sum_{g=1}^G \P[G_i = g]p_{kg} &= \sum_{g=1}^G \P[G_i = g]p_{kg}c_{kg}. \label{eq:lie_plug_in}
\end{align}

In addition to these equations, I also know that all probabilities $p_{kg}$ are non-negative and that for each $g$, $p_{1g}, \dots, p_{Kg}$ sum to 1. By Assumption \ref{ass:1}.i, any conditional expectation of $Y_i$ must be between $y_\ell$ and $y_u$.
As in \cite{cross2002regressions}, these are the only relationships I can use to relate joint information of interest to the observed marginal information in the data without any further assumptions.

I first characterize the identified set $P$ for the probability distribution of $X_i|G_i$ under Assumption \ref{ass:1} from equations \eqref{eq:lie_x_g} and \eqref{eq:lie_x} and the additional restrictions discussed above:
\begin{multline}\label{eq:ident_x}
        P = \Bigg\{(p_{11},\dots,p_{KG}) \; \Bigg\vert \; p_{kg} \geq 0 ~ \forall k,g, ~ \sum_{k=1}^K p_{kg} = 1 ~ \forall g, \\
        \text{and }\P[X_{\ell i} = x_{j,\ell}|G_i = g] = \sum_{k=1}^K \mathbbm{1}\{x_{k,\ell} = x_{j,\ell}\} p_{kg} ~ \forall g, \ell = 1,\dots,L, j=1,\dots,K \Bigg\}.
\end{multline}
Then from set $P$ and equations \eqref{eq:lie_y_g} and \eqref{eq:lie_plug_in}, I can characterize the identified set $D$ for $\sum_{k=1}^K \lambda_k \E[Y_i|X_i = x_k]$, the parameter of interest, under Assumption \ref{ass:1}:
\begin{multline}\label{eq:ident_d}
        D = \Bigg\{ \sum_{k=1}^K \lambda_k d_k \;\Bigg\vert\; \exists (p_{11}, \dots, p_{KG}) \in P, (c_{1g}, \dots, c_{KG}) \in [y_\ell,y_u]^{KG}, (d_{1}, \dots, d_{K}) \in [y_\ell,y_u]^{K} \\
        \text{s.t. } d_k \sum_{g=1}^G \P[G_i = g]p_{kg} = \sum_{g=1}^G \P[G_i = g]p_{kg}c_{kg} ~ \forall k, ~ \E[Y_i|G_i = g] = \sum_{k=1}^K c_{kg} p_{kg} ~ \forall g \Bigg\}.
\end{multline}

From this characterization it is not immediately clear how to compute the identified set. For this it is useful to re-characterize the identified set as a constrained programming problem, similar to \cite{honore2006bounds}. 
In the following proposition I show that $D$ can be written as an interval $[L,U]$, where $L$ and $U$ are the solutions to bilevel optimization problems over $\{d_k\}$, $\{c_{kg}\}$, and $\{p_{kg}\}$. Furthermore I formally state and prove that $D$ is sharp for parameter $\sum_{k=1}^K \lambda_k \E[Y_i|X_i=x_k]$.

\begin{proposition}\label{prop:program_d}
    Under Assumption \ref{ass:1}, $D$ is the sharp identified set for parameter $\sum_{k=1}^K \lambda_k \E[Y_i|X_i=x_k]$. In addition, $D$ is given by $ D = [L,U]$, where
    \begin{multline}\label{eq:opt_lower}
        L = \inf_{\{p_{kg}\},\{c_{kg}\} , \{d_k\}} \sum_{k=1}^K \lambda_k d_k \text{ s.t. } \{p_{kg}\}_{k=1,g=1}^{K,G} \in P,  
        \{c_{kg}\}_{k=1,g=1}^{K,G} \in [y_\ell,y_u]^{KG}, \{d_k\}_{k=1}^{K} \in [y_\ell, y_u]^K, \\ 
        d_k \sum_{g=1}^G \P[G_i = g]p_{kg} = \sum_{g=1}^G \P[G_i = g]p_{kg}c_{kg} ~ \forall k, \text{ and } \E[Y_i|G_i = g] = \sum_{k=1}^K c_{kg} p_{kg} ~ \forall g,
    \end{multline}
    \begin{multline}\label{eq:opt_upper}
        U = \sup_{\{p_{kg}\},\{c_{kg}\}, \{d_k\}} \sum_{k=1}^K \lambda_k d_k \text{ s.t. } \{p_{kg}\}_{k=1,g=1}^{K,G} \in P,  
        \{c_{kg}\}_{k=1,g=1}^{K,G} \in [y_\ell,y_u]^{KG}, \{d_k\}_{k=1}^{K} \in [y_\ell, y_u]^K, \\ 
        d_k \sum_{g=1}^G \P[G_i = g]p_{kg} = \sum_{g=1}^G \P[G_i = g]p_{kg}c_{kg} ~ \forall k, \text{ and } \E[Y_i|G_i = g] = \sum_{k=1}^K c_{kg} p_{kg} ~ \forall g,
    \end{multline}
    and
    \begin{multline}\label{eq:opt_p}
    P = \underset{\{p_{kg}\}}{\text{arg}\min} \sum_{g=1}^G \sum_{r=1}^{LK} v_{rg}^+ + v_{rg}^- \;\; \text{s.t. } v_{rg}^+, v_{rg}^- \geq 0 ~ \forall r,g, ~ p_{kg} \geq 0 ~ \forall k,g, ~ \sum_{k=1}^K p_{kg} = 1 ~ \forall g, \\
    \text{and } \P[X_{\ell i}=x_{k,\ell}| G_i = g] - \sum_{j=1}^K \mathbbm{1}\{x_{j,\ell} = x_{k,\ell}\} p_{jg} = v_{K(\ell-1)+k, g}^+ - v_{K(\ell-1)+k, g}^- ~ \forall \ell, k, g.
    \end{multline}
\end{proposition}
The proof of this proposition consists in showing that the characterization \eqref{eq:opt_p} is equivalent to that of \eqref{eq:ident_x} because $P$ is nonempty, and then showing that $D$ as defined in \eqref{eq:ident_d} is an interval. I defer all proof details to Appendix \ref{app:proofs}.

While the optimization problems of \eqref{eq:opt_lower} and \eqref{eq:opt_upper} have nonconvex constraints, this bilevel formulation is helpful because it suggests how computation of the lower and upper bound might be performed. In particular, the problems of \eqref{eq:opt_lower} and \eqref{eq:opt_upper} given any particular $p = \{p_{kg}\} \in P$ is a linear program in $\{c_{kg}\}$ and $\{d_k\}$. Letting $L(p)$ and $U(p)$ denote the solutions to \eqref{eq:opt_lower} and \eqref{eq:opt_upper} given $p \in P$ respectively, solving for $L(p)$ and $U(p)$ for a given $p \in P$ is a linear program and thus fast. Then I can solve for $D$ by searching for the infimum and supremum of $L(p)$ and $U(p)$, respectively, over all $p \in P$. I recommend solving for $\inf_{p \in P} L(p)$ and $\sup_{p \in P} U(p)$ using a nonconvex solver with a coarse grid of different starting points.\footnote{This may be computationally intensive when $X_i$ or $G_i$ has large support.} Checking if a candidate starting point $p$ is in $P$ is fast because the optimization problem of \eqref{eq:opt_p} is a linear program.  

\begin{remark}\label{rem:cm}
    Sharp bounds on $\E[Y_i|X_i = x_k, G_i = g]$ jointly over the support of $X_i$ can be derived under Assumption \ref{ass:1} using the method of Section 2.2 of \cite{cross2002regressions}, as discussed in Appendix \ref{app:cm}. However, I cannot use these bounds on $\E[Y_i|X_i = x_k, G_i = g]$ to directly obtain sharp bounds on (linear combinations of) $\E[Y_i|X_i = x_k]$ in my setting. 
    
    An explanation for why is that, as can be seen in \eqref{eq:lie_y_x_unknown}, $\E[Y_i|X_i = x_k]$ depends on both $\E[Y_i|X_i = x_k, G_i = g]$ and $\P[G_i = g|X_i = x_k]$. Both of these objects depend on the distribution of $X_i|G_i$, which is partially identified as set $P$ in my setting. Following \eqref{eq:lie_y_x_unknown}, one might think to obtain bounds on $\E[Y_i|X_i = x_k]$ from the set of all possible products of $\P[G_i = g|X_i = x_k]$ and $\E[Y_i|X_i = x_k, G_i = g]$, summed over $g$.
    However, the distribution of $X_i|G_i = g$ in $P$ that minimizes (maximizes) $\P[G_i = g|X_i = x_k]$ may not be the same distribution that minimizes (maximizes) $\E[Y_i|X_i = x_k, G_i = g]$ subject to \eqref{eq:lie_y_g}. 
    Thus the product of the lower (upper) bounds of the sharp identified sets for $\P[G_i = g|X_i = x_k]$ and $\E[Y_i|X_i = x_k, G_i = g]$ summed over $g$ may be strictly smaller (bigger) than the smallest (largest) possible value that $\E[Y_i|X_i = x_k]$ can take on. So this product approach does not produce sharp bounds.
    The programs of \eqref{eq:opt_lower} and \eqref{eq:opt_upper} do not have this issue because they directly minimize and maximize the parameter of interest.

    If I observe the joint distribution of $X_i|G_i$, as in \cite{cross2002regressions}, instead of each of the marginal distributions, then $\P[G_i = g|X_i = x_k]$ is point identified, as can be seen in \eqref{eq:bayes_rule}. Then I can use the sharp bounds on $\E[Y_i|X_i = x_k, G_i = g]$ derived in Appendix \ref{app:cm} from the \cite{cross2002regressions} method to simply construct bounds on linear combinations of $\E[Y_i|X_i = x_k]$ following \eqref{eq:lie_y_x_unknown}. 
    Note that if there is only a single covariate in the data, $L=1$, then I trivially observe the joint distribution of $X_i|G_i$ and thus $\P[G_i = g|X_i = x_k]$ is point-identified, as can be seen from equations \eqref{eq:lie_x_g_unknown} and \eqref{eq:bayes_rule}. This means that in the special case that $L=1$, my proposed bounds are equal to bounds on $\E[Y_i|X_i = x_k]$ obtained using the \cite{cross2002regressions} method.

    Instead of constructing an identified set for a linear combination of $\E[Y_i|X_i = x_k]$, I could also construct a (sharp) identified set for a parameter like the one considered in \cite{cross2002regressions}, namely any linear combination of $\E[Y_i|X_i = x_k, G_i = g]$, using my constrained optimization approach. In particular, using the same set $P$ in Proposition \ref{prop:program_d} I can define a new linear program that minimizes/maximizes a linear combination of $\{c_{kg}\}_{k=1,g=1}^{K,G}$ subject to the constraints that $\{c_{kg}\}_{k=1,g=1}^{K,G} \in [y_\ell, y_u]^{KG}, \{p_{kg}\}_{k=1,g=1}^{K,G} \in P$, and $\E[Y_i|G_i = g] = \sum_{k=1}^K c_{kg}p_{kg}$ for all $g$. This procedure is a bilevel constrained optimization problem where both levels are linear, for which solution methodologies have been widely studied in the optimization theory literature. The nonconvexity of the first level of the bilevel optimization program of Proposition \ref{prop:program_d} is because I target a different, albeit related, parameter.
\end{remark}

\begin{remark}\label{rem:restrictions}
    As noted by \cite{obradovic2024identification}, the formulation of the identified set as an optimization problem is convenient because additional assumptions on the data generating process can easily be accommodated by adding additional restrictions to the optimization problems, especially if the restrictions are linear or inequality restrictions. Depending on the restriction, the sharp identified set may not be an interval, but the interval defined by the lower and upper bound of the optimization problem will be the smallest closed interval containing the sharp identified set.
    
    One such restriction of interest is a polyhedral shape restriction on the conditional expectation function over groups. This can be expressed as $S_g c_{g} \leq a_g$ for each $g = 1, \dots, G$, where $c_{g} \equiv \left(c_{1g}, \dots, c_{Kg} \right)'$, $S_g \in \R^{s_g \times K}$ are known fixed matrices, and $a_g$ are known fixed vectors.
    Another restriction of interest is a homogeneity restriction on the conditional average outcome over groups, which can be expressed as $c_{kg} = c_{kg'}$ for all $k$ and $g,g'$. 
    It may also be the case that additional data at a finer level of aggregation is available to the researcher. For example, the researcher may also observe average group outcomes conditional on one of the $L$ covariates, $\E[Y_i|X_{\ell i} = x_{k,\ell},G_i=g]$. This additional data implies the restriction
    \begin{align*}
        \P[X_{\ell i} = x_{k,\ell}|G_i=g]\E[Y_i|X_{\ell i} = x_{k,\ell},G_i=g] &= \sum_{j=1}^K \mathbbm{1}\{x_{j,\ell} = x_{k,\ell}\}c_{jg}p_{jg}.
    \end{align*}
    
    I will consider the extent to which imposing these restrictions reduces the width of the identified set in the simulations and empirical illustration in Sections \ref{sec4} and \ref{sec5}. Note that if imposing a restriction makes the identified set empty, that restriction is inconsistent with the observed data. 
\end{remark}

\begin{remark}\label{rem:other_params}
    In this paper I consider parameters that are linear combinations of conditional mean outcomes $\E[Y_i|X_i]$. In principle I could target other different but related parameters that can be written in a format similar to equations \eqref{eq:lie_x_g}, \eqref{eq:lie_y_g}, and \eqref{eq:lie_plug_in} using my proposed bilevel constrained optimization approach. For example, if I observed objects like $\P[Y_i \leq y|G_i = g]$ instead of $\E[Y_i|G_i = g]$ for some fixed $y$, then I could partially identify (linear combinations of) $\P[Y_i \leq y|X_i = x_k]$ by taking $c_{kg}$ to be $\P[Y_i \leq y|X_i = x_k, G_i = g]$ and using the law of total probability statements corresponding to equations \eqref{eq:lie_y_g}, and \eqref{eq:lie_plug_in}. Such a parameter would speak to conditional quantiles of outcomes, instead of conditional mean outcomes.
    
    Another extension of my proposed method is to allow the linear combination weights $\lambda_k$ to possibly be data-dependent. If the data-dependent weights are point-identified, identification follows exactly as before and estimation can proceed using a plug-in approach similar to Section \ref{sec3_1} below. However, one would need to take into account the statistical uncertainty of these data-dependent weights when performing inference using the inference procedure of Section \ref{sec3_2}.
    I do not pursue detailed exploration of these extensions in this paper and leave them to future work.
\end{remark}

\section{Estimation and Inference}
\label{sec3}

In this section I propose a consistent estimation method for the identified set and a method for valid inference. Throughout this section I will condition on group variable $G_i$ and take it to be fixed. Thus $\P[G_i=g]$ is known without any sampling uncertainty by the researcher. Under this conditioning, the uncertainty captured by my inference method captures sampling uncertainty over individual outcomes and covariates, but not individual group assignments.

\subsection{Estimation} 
\label{sec3_1}

In practice the researcher observes sample analogs of the population values $\E[Y_i|G_i = g]$  and $\P[X_{\ell i} = x_{k,\ell} | G_i = g]$ in the aggregate data. For all $\ell = 1, \dots, L, j = 1, \dots, K, g = 1, \dots, G$, denote the observed sample analogs as
\begin{align}
    \bar{Y}_g &= \frac{\sum_{i=1}^n Y_i \mathbbm{1}\{G_i = g\}}{\sum_{i=1}^n \mathbbm{1}\{G_i = g\}} \label{eq:y_bar} \\
    P_n[X_{\ell i}=x_{j,\ell}|G_i = g] &= \frac{\sum_{i=1}^n \mathbbm{1}\{X_{\ell i}=x_{j,\ell}\} \mathbbm{1}\{G_i = g\}}{\sum_{i=1}^n \mathbbm{1}\{G_i = g\}} \label{eq:x_bar}.
\end{align}
I maintain the following assumption in order to apply a law of large numbers.\footnote{The assumption that the variables are i.i.d. can be relaxed as long as a law of large numbers still holds.}
\begin{assumption}\label{ass:5}
\begin{enumerate}[label=(\roman*)]
    \item $(Y_i, X_i): i \in \mathcal{N}_g$ are i.i.d. for each $g$, where $\mathcal{N}_g \equiv \{i \in \{1, \dots, n\}: G_i = g\}$.
    \item $\P[G_i = g] > 0$ for all $g = 1,\dots,G$.
\end{enumerate}
\end{assumption}
Under Assumption \ref{ass:5}, $\bar{Y}_g$ and $P_n[X_{\ell i}=x_{j,\ell}|G_i = g]$ both converge in probability to their respective population values as $n \to \infty$ by the law of large numbers and continuous mapping theorem, since group probabilities $\P[G_i = g]$ are positive. Thus I can construct a plug-in estimator, denoted $\hat D_n$, for the identified set by replacing all population values in the optimization problems of Proposition \ref{prop:program_d} with their sample estimates. 
The following proposition shows that the lower and upper bounds of the plug-in estimated set $\hat D_n$ are consistent.
\begin{proposition} \label{prop:consistency}
    Suppose Assumptions \ref{ass:1} and \ref{ass:5} hold. Define $\hat D_n$ with respect to Proposition \ref{prop:program_d}.
    Then the lower and upper bounds of $\hat D_n$ converge in probability to the lower and upper bounds of $D$ as $n \to \infty$.
\end{proposition}
The proof relies on Berge's maximum theorem, which requires that the set of parameters satisfying the restrictions of the optimization problem of Proposition \ref{prop:program_d} is compact-valued and continuous in the sample data. Note that both of the additional restrictions discussed in Remark \ref{rem:restrictions} are either linear restrictions or weak inequality restrictions on $\E[Y_i|X_i,G_i]$. Thus a simple extension of the proof shows that under either of the two additional restrictions, as long as the identified set is nonempty the lower and upper bounds of the estimated set are consistent.

\subsection{Inference}
\label{sec3_2}

Existing inference methods for estimated partially identified sets usually require knowledge of the joint distribution of the individual-level data to estimate a covariance matrix used in constructing critical values or test statistics for valid coverage. Examples of such methods include \cite{horowitz2000nonparametric}, \cite{imbens2004confidence}, and \cite{hsieh2022inference}, in addition to standard delta method or bootstrap approaches. However, in my setting I only observe marginal distributions of each variable, so I must consider inference methods that require only marginal information of each variable. 

I choose to construct marginal confidence intervals on each sample observation and use the Bonferroni correction to make the intervals jointly valid.
Then a valid confidence interval for the identified set is the union of the bounds from the optimization programs of Proposition \ref{prop:program_d} solved by plugging in each combination of sample observations lying within the jointly valid marginal confidence intervals. While this method produces confidence intervals that are conservative due to the Bonferroni correction, it is not immediately clear how to obtain confidence intervals that are meaningfully tighter. This approach has the advantage that constructing Bonferroni-corrected marginal confidence intervals is computationally simple, as described below. Thus to the extent that solving the programs of Proposition \ref{prop:program_d} is computationally feasible, constructing confidence intervals with this method is as well.

Sample statistics of the form $P_n[X_{\ell i} = x_{j,\ell} | G_i = g]$ are sample averages of a binary random variable, as can be seen from equation \eqref{eq:x_bar}. Thus I can construct Clopper-Pearson marginal confidence intervals, which are finite-sample valid, for each of these statistics given that I observe $n$. Note that any other binomial proportion confidence interval that obtains asymptotically nominal coverage will provide asymptotically valid coverage; Clopper-Pearson has the advantage of being finite-sample valid, even if conservative because of the finite-sample validity.

For statistics of the form $\bar{Y}_g$, to construct construct confidence intervals on the population quantities I require standard errors for these group sample means. In the case that $Y_i$ is binary, I can again construct Clopper-Pearson intervals (or any other asymptotically valid interval). 
If $Y_i$ is not binary, I can use the inequality from \cite{bhatia2000better} to bound the variance, and hence the standard errors, of the mean of each $Y_i|G_i = g$. The Bhatia-Davis inequality states that for a random variable $X$ with mean $\mu$, variance $\sigma^2$, and support with $x_\ell = \min(\supp(X)), x_u = \max(\supp(X))$,
\begin{equation*}
    \sigma^2 \leq \left(x_u - \mu \right)\left(\mu - x_\ell \right).
\end{equation*}

Thus I can estimate an upper bound for the standard error of $\bar{Y}_g$ given that I know the sample means, the support of $Y_i$, and the number of observations in each group $g$. An asymptotically valid marginal confidence interval can be constructed with the typical Gaussian limiting distribution approach, but using this upper bound instead of the standard error. Alternatively, if in the aggregate data I directly observe the standard errors for each $\bar{Y}_g$, I can construct shorter marginal confidence intervals by using the actual standard errors.

Let $M$ be the total number of statistics in the aggregate data, that is, the total number of $\bar{Y}_g$ and $P_n[X_{\ell i} = x_{k,\ell}|G_i = g]$ statistics in the data across all groups $g$, support points $k$ and covariates $\ell$.

The inference procedure is as follows:
\begin{enumerate}
    \item For every sample statistic $\hat{p}$ construct two-sided level $1-\frac{\alpha}{M}$ asymptotically valid CIs that contain $\hat p$, denoted $\left[\hat p_L, \hat p_U \right]$, as discussed above.  
    The resulting confidence intervals are 
    \begin{itemize}
        \item $\left[P_n[X_{\ell i}=x_{k,\ell}| G_i = g]_L, P_n[X_{\ell i}=x_{k,\ell}| G_i = g]_U \right]$ for each $P_n[X_{\ell i}=x_{k,\ell}| G_i = g]$,
        \item $\left[\bar{Y}_{g,L}, \bar{Y}_{g,U} \right]$ for each $\bar{Y}_g$.
    \end{itemize}
    \item Solve the optimization programs of $\hat D_n$ for all values of sample statistics within the marginal confidence intervals constructed in step 1:
        \begin{multline*}
        a) \quad \hat{P}_{CI,n} \equiv \underset{\{p_{kg}\}}{\text{arg}\min} \sum_{g=1}^G \sum_{r=1}^{2LK} v_{rg}^+ + v_{rg}^- \;\; \text{s.t. } p_{kg}, v_{rg}^+, v_{rg}^- \geq 0 ~ \forall k, r, g, \; \sum_{k=1}^K p_{kg} = 1 ~ \forall g, \\
        P_n[X_{\ell i}=x_{k,\ell}| G_i = g]_L - \sum_{k=1}^K \mathbbm{1}\{x_{j,\ell} = x_{k,\ell}\} p_{jg} \leq \; v_{K(\ell-1)+k}^+ - v_{K(\ell-1)+k}^- ~ \forall \ell, k, g, \text{ and }\\
        P_n[X_{\ell i}=x_{k,\ell}| G_i = g]_U - \sum_{k=1}^K \mathbbm{1}\{x_{j,\ell} = x_{k,\ell}\} p_{jg} \geq v_{K(L+\ell-1)+k}^+ - v_{K(L+\ell-1)+k}^- ~ \forall \ell, k, g.
        \end{multline*}
        \vspace{-20pt}
        \begin{multline*}
        b) \quad \hat L_{CI,n} \equiv \inf_{\{p_{kg}\},\{c_{kg}\}, \{d_k\}} \sum_{k=1}^K \lambda_k d_k  \text{ s.t. } \{p_{kg}\}_{k=1,g=1}^{K,G} \in \hat P_{CI,n}, \{c_{kg}\}_{k=1,g=1}^{K,G} \in [y_\ell,y_u]^{KG}, \\
        \{d_k\}_{k=1}^{K} \in [y_\ell, y_u]^K, \;
        \bar{Y}_{g,L} \leq \sum_{k=1}^K c_{kg}p_{kg} ~ \forall g, \; \bar{Y}_{g,U} \geq \sum_{k=1}^K c_{kg}p_{kg} ~ \forall g, \\
        \text{and } d_k \sum_{g=1}^G \P[G_i = g] p_{kg} = \sum_{g=1}^G \P[G_i = g] p_{kg}c_{kg} ~ \forall k.
        \end{multline*}
        \vspace{-20pt}
        \begin{multline*} 
        c) \quad \hat U_{CI,n} \equiv \sup_{\{p_{kg}\},\{c_{kg}\}, \{d_k\}} \sum_{k=1}^K \lambda_k d_k  \text{ s.t. } \{p_{kg}\}_{k=1,g=1}^{K,G} \in \hat P_{CI,n}, \{c_{kg}\}_{k=1,g=1}^{K,G} \in [y_\ell,y_u]^{KG}, \\
        \{d_k\}_{k=1}^{K} \in [y_\ell, y_u]^K, \;
        \bar{Y}_{g,L} \leq \sum_{k=1}^K c_{kg}p_{kg} ~ \forall g, \; \bar{Y}_{g,U} \geq \sum_{k=1}^K c_{kg}p_{kg} ~ \forall g, \\
        \text{and } d_k \sum_{g=1}^G \P[G_i = g] p_{kg} = \sum_{g=1}^G \P[G_i = g] p_{kg}c_{kg} ~ \forall k.
        \end{multline*}
    \item The confidence interval is given by $\hat D_{CI,n} \equiv [\hat L_{CI,n}, \hat U_{CI,n}]$.
\end{enumerate}

\begin{proposition} \label{prop:inference}
    Suppose Assumption \ref{ass:1} holds. Define $\hat D_{CI,n}$ as in the inference procedure described above. 
    Then $\displaystyle \lim_{n \to \infty} \P[D \subseteq \hat D_{CI,n}] \geq 1-\alpha$.
\end{proposition}
Proposition \ref{prop:inference} says that the confidence interval $\hat D_{CI,n}$ has correct asymptotic coverage for the identified set. Since $\sum_{k=1}^K \lambda_k \E[Y_i|X_i = x_k] \in D$, $ \lim_{n \to \infty} \P\left[\sum_{k=1}^K \lambda_k \E [Y_i|X_i = x_k] \in \hat D_{CI,n} \right] \geq 1-\alpha$, and thus the confidence interval also has correct asymptotic coverage for the target parameter, as discussed in \cite{imbens2004confidence}. Note that if outcome $Y_i$ is binary and one uses Clopper-Pearson confidence intervals in step 1 for all statistics, coverage in the above proposition is finite-sample valid.

\section{Simulations}
\label{sec4}

In this section I present results from three simulation studies in order to illustrate the coverage properties of the inference procedure of Section \ref{sec3_2} and what features of the aggregate data drive the width of the identified set. I also compare the sharp bounds obtained using my method to bounds obtained using the product approach with the \cite{cross2002regressions} method in Remark \ref{rem:cm}.

I consider a binary outcome $Y_i \in \{0,1\}$, three binary covariates $X_i = (X_{1i}, X_{2i}, X_{3i}) \in \{0,1\}^3$, and five groups $G_i \in \{1, \dots, 5\}$. Across all three simulation studies I assume the following model for how latent individual outcome $Y_i$ relates to latent covariates $X_i$:
\begin{align*}
   Y_i = \mathbbm{1}\{4X_{1i} - 9X_{2i} - 4X_{3i} - 1 \geq u_i\}, \qquad u_i \overset{i.i.d.}{\sim} N(0,1).
\end{align*}
I present the true population aggregate features of the data for the first simulation study in Table \ref{tab:sim_true_1}, for the second simulation study in Table \ref{tab:sim_true_2}, and for the third simulation study in Table \ref{tab:sim_true_3}. In each study $\E[Y_i|G_i]$ is implied by the choice of marginal distributions for covariates $X_i$ and the above model. The data-generating process of the first study is calibrated to approximate the data used in the empirical illustration. In the second study, I maintain the same marginal distribution for $X_{3i}$ but push the distributions for $X_{1i}$ and $X_{2i}$ much closer to 1 and 0 respectively. In the third study I maintain the same marginal distribution for $X_{2i}$ and $X_{3i}$ as in the second study, but instead enforce that some groups have almost all individuals with $X_{1i} = 0$ and other groups have almost all individuals with $X_{1i} = 1$.

\begin{table}[!htbp] 
    \caption{True population aggregate features of data for first simulation study}
    \label{tab:sim_true_1}
    \centering
    \resizebox{0.8\textwidth}{!}{
    \begin{tabular}{ccccccc}
    \hline\hline 
    & & \\[-12pt]
    $G_i$ & $\E[Y_i|G_i]$ & $\P[X_{1i}=1|G_i]$ & $\P[X_{2i}=1|G_i]$ & $\P[X_{3i}=1|G_i]$ & $\P[G_i=g]$ \\
    \hline
    1 & 0.246 & 0.389 & 0.447 & 0.096 & 0.625 \\[5pt]
    2 & 0.560 & 0.801 & 0.273 & 0.087 & 0.094 \\[5pt]
    3 & 0.610 & 0.811 & 0.233 & 0.064 & 0.094 \\[5pt]
    4 & 0.729 & 0.859 & 0.149 & 0.031 & 0.094 \\[5pt]
    5 & 0.346 & 0.900 & 0.620 & 0.004 & 0.094 \\
    \hline\hline \\[-10pt]
    \end{tabular}}
    \begin{minipage}{\textwidth}
    \fontsize{9}{2}\linespread{1}\selectfont
    \footnotesize{
    \textit{Notes}: See Section \ref{sec4} for more details on how the data set is constructed. All numbers are rounded to the nearest thousandth.
    }
    \end{minipage}
\end{table}

\begin{table}[!htbp] 
    \caption{True population aggregate features of data for second simulation study}
    \label{tab:sim_true_2}
    \centering
    \resizebox{0.8\textwidth}{!}{
    \begin{tabular}{ccccccc}
    \hline\hline 
    & & \\[-12pt]
    $G_i$ & $\E[Y_i|G_i]$ & $\P[X_{1i}=1|G_i]$ & $\P[X_{2i}=1|G_i]$ & $\P[X_{3i}=1|G_i]$ & $\P[G_i=g]$ \\
    \hline
    1 & 0.553 & 0.732 & 0.194 & 0.096 & 0.625 \\[5pt]
    2 & 0.712 & 0.893 & 0.126 & 0.087 & 0.094 \\[5pt]
    3 & 0.744 & 0.896 & 0.111 & 0.064 & 0.094 \\[5pt]
    4 & 0.810 & 0.914 & 0.079 & 0.031 & 0.094 \\[5pt]
    5 & 0.744 & 0.931 & 0.189 & 0.004 & 0.094 \\
    \hline\hline\\[-10pt]
    \end{tabular}}
    \begin{minipage}{\textwidth}
    \fontsize{9}{2}\linespread{1}\selectfont
    \footnotesize{
    \textit{Notes}: See Section \ref{sec4} for more details on how the data set is constructed. All numbers are rounded to the nearest thousandth.
    }
    \end{minipage}
\end{table}

\begin{table}[!htbp] 
    \caption{True population aggregate features of data for third simulation study}
    \label{tab:sim_true_3}
    \centering
    \resizebox{0.8\textwidth}{!}{
    \begin{tabular}{ccccccc}
    \hline\hline 
    & & \\[-12pt]
    $G_i$ & $\E[Y_i|G_i]$ & $\P[X_{1i}=1|G_i]$ & $\P[X_{2i}=1|G_i]$ & $\P[X_{3i}=1|G_i]$ & $\P[G_i=g]$ \\
    \hline
    1 & 0.122 & 0.010 & 0.194 & 0.096 & 0.625 \\[5pt]
    2 & 0.187 & 0.092 & 0.126 & 0.087 & 0.094 \\[5pt]
    3 & 0.254 & 0.179 & 0.111 & 0.064 & 0.094 \\[5pt]
    4 & 0.842 & 0.958 & 0.079 & 0.031 & 0.094 \\[5pt]
    5 & 0.774 & 0.977 & 0.189 & 0.004 & 0.094 \\
    \hline\hline\\[-10pt]
    \end{tabular}}
    \begin{minipage}{\textwidth}
    \fontsize{9}{2}\linespread{1}\selectfont
    \footnotesize{
    \textit{Notes}: See Section \ref{sec4} for more details on how the data set is constructed. All numbers are rounded to the nearest thousandth.
    }
    \end{minipage}
\end{table}

In accordance with the sampling thought experiment used for inference in Section \ref{sec3}, in which I condition on group assignment $G_i$ and thus take it to be fixed, I create 500 aggregate data sets from each of the three true data-generating processes by, for each individual in group $g$, randomly drawing $X_{1i}, X_{2i},$ and $X_{3i}$ each according to the true marginal distributions for $X_{1i}|G_i=g, X_{2i}|G_i=g$, and $X_{3i}|G_i=g$ respectively. I then compute $Y_i$ according to the model above. Once $(X_i, Y_i)$ is drawn for all individuals in group $g$, I aggregate variables across individuals in each group to create a sample aggregate data set. Note that while I draw each sample $X_i$ as if $X_{1i}, X_{2i}$, and $X_{3i}$ are independent within each group, this is without loss of generality for the simulation because the covariance structure of $X_i$ within each group does not matter for the aggregate sample data set.

I consider two kinds of parameters across the simulation studies: conditional mean outcomes $\E[Y_i|X_i]$ and conditional outcome differences across $X_{1i}$, $\E[Y_i|X_{1i}=1,X_{2i},X_{3i}] - \E[Y_i|X_{1i}=0,X_{2i},X_{3i}]$. These are the same as the parameters that I consider in the empirical illustration. For each simulation study and parameter of interest I compute the population identified set using the population aggregate features without any additional restrictions and then with the monotonicity restrictions used later in the empirical illustration:
\begin{equation}
\begin{gathered} \label{eq:sims_mono}
    \E[Y_i|X_{2i}=1,X_{1i},X_{3i},G_i=g] - \E[Y_i|X_{2i}=0,X_{1i},X_{3i},G_i=g] \leq 0 \\
    \E[Y_i|X_{3i}=1,X_{1i},X_{2i},G_i=g] - \E[Y_i|X_{3i}=0,X_{1i},X_{2i},G_i=g] \leq 0.
\end{gathered}
\end{equation}
I then compute estimated bounds and 90\% confidence intervals for each of the 500 sample data sets in each simulation study.

In Table \ref{tab:sim_results_1} I report results from the first simulation study. In particular, I report the population identified set, average coverage of the 90\% confidence intervals, the ratio of the average width of the estimated bounds to the width of the identified set, and the ratio of the average width of the 90\% confidence intervals to the width of the identified set, both without restrictions and under monotonicity. Bounds are not very informative across all parameters, although monotonicity does have some identifying power for certain parameters. As expected, coverage is conservative for all parameters. However, the average width of the confidence intervals is no more than 3\% more than the width of the identified set. Similarly, the average width of the estimated bounds is essentially the same as the width of the identified set.

\begin{table}[!htbp] 
    \caption{Identified sets and coverage from first simulation study}
    \label{tab:sim_results_1}
    \centering
    \resizebox{\textwidth}{!}{
    \begin{tabular}{ccccccccc}
    \hline\hline 
    & & \\[-12pt]
    & \multicolumn{4}{c}{\textbf{No monotonicity}} & \multicolumn{4}{c}{\textbf{Monotonicity}} \\
    \cmidrule(lr){2-5} \cmidrule(lr){6-9}
     & Identified &  & Relative & Relative & Identified & & Relative & Relative \\
     & set & Coverage & est. width & CI width & set & Coverage & est. width & CI width \\
    Parameter & (1) & (2) & (3) & (4) & (5) & (6) & (7) & (8) \\
    \hline
    $\E[Y_i|X_{1i}=1,X_{2i}=0,X_{3i}=0]$ & \multirow{2}{*}{$[-0.884,1]$} & \multirow{2}{*}{1} & \multirow{2}{*}{0.999} & \multirow{2}{*}{1.012} & \multirow{2}{*}{$[-0.777,1]$} & \multirow{2}{*}{1} & \multirow{2}{*}{1.000} & \multirow{2}{*}{1.013}\\
    $-\E[Y_i|X_{1i}=0,X_{2i}=0,X_{3i}=0]$ &  &  &  &  \\[10pt]
    $\E[Y_i|X_{1i}=1,X_{2i}=1,X_{3i}=0]$ & \multirow{2}{*}{$[-1,0.940]$} & \multirow{2}{*}{1} & \multirow{2}{*}{1.000} & \multirow{2}{*}{1.010} & \multirow{2}{*}{$[-1, 0.760]$} & \multirow{2}{*}{1} & \multirow{2}{*}{1.000} & \multirow{2}{*}{1.029} \\
    $-\E[Y_i|X_{1i}=0,X_{2i}=1,X_{3i}=0]$ &  &  &  &  \\[10pt]
    $\E[Y_i|X_{1i}=1,X_{2i}=0,X_{3i}=1]$ & \multirow{2}{*}{$[-1,1]$} & \multirow{2}{*}{1} & \multirow{2}{*}{1} & \multirow{2}{*}{1} & \multirow{2}{*}{$[-1,1]$} & \multirow{2}{*}{1} & \multirow{2}{*}{1} & \multirow{2}{*}{1} \\
    $-\E[Y_i|X_{1i}=0,X_{2i}=0,X_{3i}=1]$ &  &  &  &  \\[10pt]
    $\E[Y_i|X_{1i}=1,X_{2i}=1,X_{3i}=1]$ & \multirow{2}{*}{$[-1,1]$} & \multirow{2}{*}{1} & \multirow{2}{*}{1} & \multirow{2}{*}{1} & \multirow{2}{*}{$[-1,1]$} & \multirow{2}{*}{1} & \multirow{2}{*}{1} & \multirow{2}{*}{1} \\
    $-\E[Y_i|X_{1i}=0,X_{2i}=1,X_{3i}=1]$ &  &  &  &  \\[10pt]
    $\E[Y_i|X_{1i}=0,X_{2i}=0,X_{3i}=0]$ & $[0,1]$ & 1 & 1 & 1 & $[0,1]$ & 1 & 1 & 1 \\[10pt]
    $\E[Y_i|X_{1i}=1,X_{2i}=0,X_{3i}=0]$ & $[0.116,1]$ & 1 & 0.998 & 1.027 & $[0.223,1]$ & 1 & 1.000 & 1.029 \\[10pt]
    $\E[Y_i|X_{1i}=0,X_{2i}=1,X_{3i}=0]$ & $[0,1]$ & 1 & 1 & 1 & $[0,1]$ & 1 & 1 & 1 \\[10pt]
    $\E[Y_i|X_{1i}=0,X_{2i}=0,X_{3i}=1]$ & $[0,1]$ & 1 & 1 & 1 & $[0,1]$ & 1 & 1 & 1 \\[10pt]
    $\E[Y_i|X_{1i}=1,X_{2i}=1,X_{3i}=0]$ & $[0,0.940]$ & 1 & 1.000 & 1.020 & $[0,0.760]$ & 1 & 1.000 & 1.068 \\[10pt]
    $\E[Y_i|X_{1i}=1,X_{2i}=0,X_{3i}=1]$ & $[0,1]$ & 1 & 1 & 1 & $[0,1]$ & 1 & 1 & 1 \\[10pt]
    $\E[Y_i|X_{1i}=0,X_{2i}=1,X_{3i}=1]$ & $[0,1]$ & 1 & 1 & 1 & $[0,1]$ & 1 & 1 & 1 \\[10pt]
    $\E[Y_i|X_{1i}=1,X_{2i}=1,X_{3i}=1]$ & $[0,1]$ & 1 & 1 & 1 & $[0,1]$ & 1 & 1 & 1 \\
    \hline\hline\\[-10pt]
    \end{tabular}}
    \begin{minipage}{\textwidth}
    \fontsize{9}{2}\linespread{1}\selectfont
    \footnotesize{
    \textit{Notes}: Reported bounds are population identified sets $D = [L,U]$ under no additional restrictions following the program of Proposition \ref{prop:program_d} (column 1) and the additional monotonicity restriction of \eqref{eq:sims_mono} (column 5). 90\% confidence intervals for each of the 500 sample data sets are constructed following the inference procedure of Section \ref{sec3_2} and average coverage of the identified set is reported in column 2 without restrictions and column 6 under additional monotonicity restrictions. 
    The relative estimated width, reported in columns 3 and 7, is the average of the width of the estimated bounds from each of the 500 sample data sets, divided by the width of the identified set. Similarly, the relative CI width, reported in columns 4 and 8, is the average of the width of the estimated 90\% confidence intervals from each of the 500 sample data sets, divided by the width of the identified set.
    All numbers are rounded to the nearest thousandth. Relative widths of 1.000 are rounded while relative widths of 1 are exactly 1.
    }
    \end{minipage}
\end{table}

In Table \ref{tab:sim_results_2} I report results from the second simulation study analogous to those of Table \ref{tab:sim_results_1}.
Coverage is again conservative, although confidence intervals are not too much wider than the identified set on average, and estimated bounds are the same width as the identified set on average. 
Note from Table \ref{tab:sim_true_2} that the data is most representative of groups with high proportions of $X_{1i}$ and low proportions of $X_{2i}$ and $X_{3i}$. Bounds on the parameters corresponding to this subgroup---$\E[Y_i|X_{1i}=1,X_{2i}=0,X_{3i}=0]$ and $\E[Y_i|X_{1i}=1,X_{2i}=0,X_{3i}=0] - \E[Y_i|X_{1i}=0,X_{2i}=0,X_{3i}=0]$---are more informative in the second study compared to the first simulation study. Bounds on all other parameters are weakly less informative than the first simulation study.

This suggests that a feature of the aggregate data that shortens the width of bounds on conditional outcome parameters involving $X_i = x_k$ is how well-represented individuals with $X_{\ell i} = x_{k,\ell}$ for each covariate $\ell$ are in the data. For example, bounds on parameters involving $X_i = (1,0,0)$ are more informative in the second study than the first study because more individuals in the second study have each of $X_{1i} = 1, X_{2i} = 0,$ and $X_{3i} = 0$.
Theoretically this feature helps because if marginal probabilities $\P[X_{\ell i} = x_{k,\ell}|G_i = g]$ are close to 1 across all $\ell$, there are fewer possible values the joint probability $\P[X_i=x_k|G_i=g]$ can take on. This reduces the width of the identified set for $\mathbb{P}[X_i=x_k|G_i=g]$ which then helps to narrow the bounds, as can be seen from the bilevel program representation of the identified set.

\begin{table}[!htbp] 
    \caption{Identified sets and coverage from second simulation study}
    \label{tab:sim_results_2}
    \centering
    \resizebox{\textwidth}{!}{
    \begin{tabular}{ccccccccc}
    \hline\hline 
    & & \\[-12pt]
    & \multicolumn{4}{c}{\textbf{No monotonicity}} & \multicolumn{4}{c}{\textbf{Monotonicity}} \\
    \cmidrule(lr){2-5} \cmidrule(lr){6-9}
     & Identified &  & Relative & Relative & Identified & & Relative & Relative \\
     & set & Coverage & est. width & CI width & set & Coverage & est. width & CI width \\
    Parameter & (1) & (2) & (3) & (4) & (5) & (6) & (7) & (8) \\
    \hline
    $\E[Y_i|X_{1i}=1,X_{2i}=0,X_{3i}=0]$ & \multirow{2}{*}{$[-0.670,1]$} & \multirow{2}{*}{1} & \multirow{2}{*}{1.000} & \multirow{2}{*}{1.017} & \multirow{2}{*}{$[-0.482,1]$} & \multirow{2}{*}{1} & \multirow{2}{*}{1.000} & \multirow{2}{*}{1.013} \\
    $-\E[Y_i|X_{1i}=0,X_{2i}=0,X_{3i}=0]$ &  &  &  &  \\[10pt]
    $\E[Y_i|X_{1i}=1,X_{2i}=1,X_{3i}=0]$ & \multirow{2}{*}{$[-1,1]$} & \multirow{2}{*}{1} & \multirow{2}{*}{1} & \multirow{2}{*}{1} & \multirow{2}{*}{$[-1, 0.871]$} & \multirow{2}{*}{1} & \multirow{2}{*}{1.001} & \multirow{2}{*}{1.024} \\
    $-\E[Y_i|X_{1i}=0,X_{2i}=1,X_{3i}=0]$ &  &  &  &  \\[10pt]
    $\E[Y_i|X_{1i}=1,X_{2i}=0,X_{3i}=1]$ & \multirow{2}{*}{$[-1,1]$} & \multirow{2}{*}{1} & \multirow{2}{*}{1} & \multirow{2}{*}{1} & \multirow{2}{*}{$[-1,1]$} & \multirow{2}{*}{1} & \multirow{2}{*}{1} & \multirow{2}{*}{1} \\
    $-\E[Y_i|X_{1i}=0,X_{2i}=0,X_{3i}=1]$ &  &  &  &  \\[10pt]
    $\E[Y_i|X_{1i}=1,X_{2i}=1,X_{3i}=1]$ & \multirow{2}{*}{$[-1,1]$} & \multirow{2}{*}{1} & \multirow{2}{*}{1} & \multirow{2}{*}{1} & \multirow{2}{*}{$[-1,1]$} & \multirow{2}{*}{1} & \multirow{2}{*}{1} & \multirow{2}{*}{1} \\
    $-\E[Y_i|X_{1i}=0,X_{2i}=1,X_{3i}=1]$ &  &  &  &  \\[10pt]
    $\E[Y_i|X_{1i}=0,X_{2i}=0,X_{3i}=0]$ & $[0,1]$ & 1 & 1 & 1 & $[0,1]$ & 1 & 1 & 1 \\[10pt]
    $\E[Y_i|X_{1i}=1,X_{2i}=0,X_{3i}=0]$ & $[0.330,1]$ & 1 & 1.000 & 1.041 & $[0.518,1]$ & 1 & 1.000 & 1.041 \\[10pt]
    $\E[Y_i|X_{1i}=0,X_{2i}=1,X_{3i}=0]$ & $[0,1]$ & 1 & 1 & 1 & $[0,1]$ & 1 & 1 & 1 \\[10pt]
    $\E[Y_i|X_{1i}=0,X_{2i}=0,X_{3i}=1]$ & $[0,1]$ & 1 & 1 & 1 & $[0,1]$ & 1 & 1 & 1 \\[10pt]
    $\E[Y_i|X_{1i}=1,X_{2i}=1,X_{3i}=0]$ & $[0,1]$ & 1 & 1 & 1 & $[0,0.871]$ & 1 & 1.001 & 1.052 \\[10pt]
    $\E[Y_i|X_{1i}=1,X_{2i}=0,X_{3i}=1]$ & $[0,1]$ & 1 & 1 & 1 & $[0,1]$ & 1 & 1 & 1 \\[10pt]
    $\E[Y_i|X_{1i}=0,X_{2i}=1,X_{3i}=1]$ & $[0,1]$ & 1 & 1 & 1 & $[0,1]$ & 1 & 1 & 1 \\[10pt]
    $\E[Y_i|X_{1i}=1,X_{2i}=1,X_{3i}=1]$ & $[0,1]$ & 1 & 1 & 1 & $[0,1]$ & 1 & 1 & 1 \\
    \hline\hline\\[-10pt]
    \end{tabular}}
    \begin{minipage}{\textwidth}
    \fontsize{9}{2}\linespread{1}\selectfont
    \footnotesize{
    \textit{Notes}: Reported bounds are population identified sets $D = [L,U]$ under no additional restrictions following the program of Proposition \ref{prop:program_d} (column 1) and the additional monotonicity restriction of \eqref{eq:sims_mono} (column 5). 90\% confidence intervals for each of the 500 sample data sets are constructed following the inference procedure of Section \ref{sec3_2} and average coverage of the identified set is reported in column 2 without restrictions and column 6 under additional monotonicity restrictions. 
    The relative estimated width, reported in columns 3 and 7, is the average of the width of the estimated bounds from each of the 500 sample data sets, divided by the width of the identified set. Similarly, the relative CI width, reported in columns 4 and 8, is the average of the width of the estimated 90\% confidence intervals from each of the 500 sample data sets, divided by the width of the identified set.
    All numbers are rounded to the nearest thousandth. Relative widths of 1.000 are rounded while relative widths of 1 are exactly 1.
    }
    \end{minipage}
\end{table}

In Table \ref{tab:sim_results_3} I report results analogous to those above for the third simulation study. The third simulation study demonstrates the potential of my partial identification method to recover informative bounds. For example, bounds on the conditional outcome difference parameter $\E[Y_i|X_{1i}=1,X_{2i}=0,X_{3i}=0]-\E[Y_i|X_{1i}=0,X_{2i}=0,X_{3i}=0]$ do not contain zero. As before coverage is conservative although confidence intervals are not too much wider than the identified set on average, while estimated bounds are the same width as the identified set on average.

In this simulation study the data is representative of groups with both high and low proportions of $X_{1i}$, but only low proportions of $X_{2i}$ and $X_{3i}$, where high proportions of $X_{1i}$ occur with high expected outcome $Y_i$ and low proportions of $X_{1i}$ occur with low expected $Y_i$. As suggested by the results of the second simulation study, the parameters corresponding to these subgroups---$\E[Y_i|X_{1i}=0,X_{2i}=0,X_{3i}=0], \E[Y_i|X_{1i}=1,X_{2i}=0,X_{3i}=0]$, and $\E[Y_i|X_{1i}=1,X_{2i}=0,X_{3i}=0] - \E[Y_i|X_{1i}=0,X_{2i}=0,X_{3i}=0]$---have bounds that are much more informative relative to both the first and second simulation studies.

\begin{table}[!htbp] 
    \caption{Identified sets and coverage from third simulation study}
    \label{tab:sim_results_3}
    \centering
    \resizebox{\textwidth}{!}{
    \begin{tabular}{ccccccccc}
    \hline\hline 
    & & \\[-12pt]
    & \multicolumn{4}{c}{\textbf{No monotonicity}} & \multicolumn{4}{c}{\textbf{Monotonicity}} \\
    \cmidrule(lr){2-5} \cmidrule(lr){6-9}
     & Identified &  & Relative & Relative & Identified & & Relative & Relative \\
     & set & Coverage & est. width & CI width & set & Coverage & est. width & CI width \\
    Parameter & (1) & (2) & (3) & (4) & (5) & (6) & (7) & (8) \\
    \hline
    $\E[Y_i|X_{1i}=1,X_{2i}=0,X_{3i}=0]$ & \multirow{2}{*}{$[0.417,0.992]$} & \multirow{2}{*}{1} & \multirow{2}{*}{1.000} & \multirow{2}{*}{1.105} & \multirow{2}{*}{$[0.448,0.884]$} & \multirow{2}{*}{1} & \multirow{2}{*}{1.000} & \multirow{2}{*}{1.154} \\
    $-\E[Y_i|X_{1i}=0,X_{2i}=0,X_{3i}=0]$ &  &  &  &  \\[10pt]
    $\E[Y_i|X_{1i}=1,X_{2i}=1,X_{3i}=0]$ & \multirow{2}{*}{$[-1,1]$} & \multirow{2}{*}{1} & \multirow{2}{*}{1} & \multirow{2}{*}{1} & \multirow{2}{*}{$[-0.239, 0.923]$} & \multirow{2}{*}{1} & \multirow{2}{*}{1.000} & \multirow{2}{*}{1.049} \\
    $-\E[Y_i|X_{1i}=0,X_{2i}=1,X_{3i}=0]$ &  &  &  &  \\[10pt]
    $\E[Y_i|X_{1i}=1,X_{2i}=0,X_{3i}=1]$ & \multirow{2}{*}{$[-1,1]$} & \multirow{2}{*}{1} & \multirow{2}{*}{1} & \multirow{2}{*}{1} & \multirow{2}{*}{$[-1,1]$} & \multirow{2}{*}{1} & \multirow{2}{*}{1} & \multirow{2}{*}{1} \\
    $-\E[Y_i|X_{1i}=0,X_{2i}=0,X_{3i}=1]$ &  &  &  &  \\[10pt]
    $\E[Y_i|X_{1i}=1,X_{2i}=1,X_{3i}=1]$ & \multirow{2}{*}{$[-1,1]$} & \multirow{2}{*}{1} & \multirow{2}{*}{1} & \multirow{2}{*}{1} & \multirow{2}{*}{$[-1,1]$} & \multirow{2}{*}{1} & \multirow{2}{*}{1} & \multirow{2}{*}{1} \\
    $-\E[Y_i|X_{1i}=0,X_{2i}=1,X_{3i}=1]$ &  &  &  &  \\[10pt]
    $\E[Y_i|X_{1i}=0,X_{2i}=0,X_{3i}=0]$ & $[0,0.217]$ & 1 & 1.000 & 1.080 & $[0.108,0.217]$ & 1 & 1.000 & 1.277 \\[10pt]
    $\E[Y_i|X_{1i}=1,X_{2i}=0,X_{3i}=0]$ & $[0.633,0.992]$ & 1 & 1.000 & 1.121 & $[0.665,0.992]$ & 1 & 1.000 & 1.112 \\[10pt]
    $\E[Y_i|X_{1i}=0,X_{2i}=1,X_{3i}=0]$ & $[0,1]$ & 1 & 1 & 1 & $[0,0.239]$ & 1 & 1.000 & 1.115 \\[10pt]
    $\E[Y_i|X_{1i}=0,X_{2i}=0,X_{3i}=1]$ & $[0,1]$ & 1 & 1 & 1 & $[0,1]$ & 1 & 1 & 1 \\[10pt]
    $\E[Y_i|X_{1i}=1,X_{2i}=1,X_{3i}=0]$ & $[0,1]$ & 1 & 1 & 1 & $[0,0.923]$ & 1 & 1.000 & 1.021 \\[10pt]
    $\E[Y_i|X_{1i}=1,X_{2i}=0,X_{3i}=1]$ & $[0,1]$ & 1 & 1 & 1 & $[0,1]$ & 1 & 1 & 1 \\[10pt]
    $\E[Y_i|X_{1i}=0,X_{2i}=1,X_{3i}=1]$ & $[0,1]$ & 1 & 1 & 1 & $[0,1]$ & 1 & 1 & 1 \\[10pt]
    $\E[Y_i|X_{1i}=1,X_{2i}=1,X_{3i}=1]$ & $[0,1]$ & 1 & 1 & 1 & $[0,1]$ & 1 & 1 & 1  \\
    \hline\hline\\[-10pt]
    \end{tabular}}
    \begin{minipage}{\textwidth}
    \fontsize{9}{2}\linespread{1}\selectfont
    \footnotesize{
    \textit{Notes}: Reported bounds are population identified sets $D = [L,U]$ under no additional restrictions following the program of Proposition \ref{prop:program_d} (column 1) and the additional monotonicity restriction of \eqref{eq:sims_mono} (column 5). 90\% confidence intervals for each of the 500 sample data sets are constructed following the inference procedure of Section \ref{sec3_2} and average coverage of the identified set is reported in column 2 without restrictions and column 6 under additional monotonicity restrictions. 
    The relative estimated width, reported in columns 3 and 7, is the average of the width of the estimated bounds from each of the 500 sample data sets, divided by the width of the identified set. Similarly, the relative CI width, reported in columns 4 and 8, is the average of the width of the estimated 90\% confidence intervals from each of the 500 sample data sets, divided by the width of the identified set.
    All numbers are rounded to the nearest thousandth. Relative widths of 1.000 are rounded while relative widths of 1 are exactly 1.
    }
    \end{minipage}
\end{table}

\paragraph{Comparison with \cite{cross2002regressions} bounds}
In Appendix \ref{app:cm} I show how to obtain sharp bounds on $\E[Y_i|X_i=x_k,G_i=g]$ with aggregate data using the method proposed in \cite{cross2002regressions}. I note in Remark \ref{rem:cm} that although one can use the \cite{cross2002regressions} approach on $\E[Y_i|X_i=x_k,G_i=g]$ together with the identified set $P$ for the distribution of $X_i|G_i$ to obtain valid bounds on linear combinations of $\E[Y_i|X_i=x_k]$, these bounds are not sharp. Note that this alternative approach is also a nonconvex optimization problem for similar reasons that the approach I propose is nonconvex.

I demonstrate this numerically in Table \ref{tab:cm} for the three aggregate data sets used in the simulation studies. For all parameters considered in the simulation study, I display the sharp identified set obtained from my method (columns 1, 3, and 5) and the bounds obtained from following the approach described in Remark \ref{rem:cm} using \cite{cross2002regressions} bounds. The sharp identified set is contained in the bound from the \cite{cross2002regressions} approach for all parameters. In fact, for all but one of the parameters for which the identified set is strictly informative, the sharp identified set is a strict subset of the bound from the \cite{cross2002regressions} approach.

\begin{table}[!htbp] 
    \caption{Identified sets under my method and \cite{cross2002regressions} method}
    \label{tab:cm}
    \centering
    \resizebox{\textwidth}{!}{
    \begin{tabular}{ccccccc}
    \hline\hline 
    & & \\[-12pt]
    & \multicolumn{2}{c}{\textbf{First study}} & \multicolumn{2}{c}{\textbf{Second study}} & \multicolumn{2}{c}{\textbf{Third study}} \\
    \cmidrule(lr){2-3} \cmidrule(lr){4-5} \cmidrule(lr){6-7}
      & Identified set & CM set & Identified set & CM set & Identified set & CM set \\
    Parameter & (1) & (2) & (3) & (4) & (5) & (6) \\
    \hline
    $\E[Y_i|X_{1i}=1,X_{2i}=0,X_{3i}=0]$ & \multirow{2}{*}{$[-0.884,1]$} & \multirow{2}{*}{$[-0.891, 1]$} & \multirow{2}{*}{$[-0.670,1]$} & \multirow{2}{*}{$[-0.750,1]$} & \multirow{2}{*}{$[0.417,0.992]$} & \multirow{2}{*}{$[0.414,0.992]$} \\
    $-\E[Y_i|X_{1i}=0,X_{2i}=0,X_{3i}=0]$ &  &  \\[10pt]
    $\E[Y_i|X_{1i}=1,X_{2i}=1,X_{3i}=0]$ & \multirow{2}{*}{$[-1,0.940]$} & \multirow{2}{*}{$[-1,0.955]$} & \multirow{2}{*}{$[-1,1]$} & \multirow{2}{*}{$[-1,1]$} & \multirow{2}{*}{$[-1,1]$} & \multirow{2}{*}{$[-1,1]$} \\
    $-\E[Y_i|X_{1i}=0,X_{2i}=1,X_{3i}=0]$ &  & \\[10pt]
    $\E[Y_i|X_{1i}=1,X_{2i}=0,X_{3i}=1]$ & \multirow{2}{*}{$[-1,1]$} & \multirow{2}{*}{$[-1,1]$} & \multirow{2}{*}{$[-1,1]$} & \multirow{2}{*}{$[-1,1]$} & \multirow{2}{*}{$[-1,1]$} & \multirow{2}{*}{$[-1,1]$} \\
    $-\E[Y_i|X_{1i}=0,X_{2i}=0,X_{3i}=1]$ &  & \\[10pt]
    $\E[Y_i|X_{1i}=1,X_{2i}=1,X_{3i}=1]$ & \multirow{2}{*}{$[-1,1]$} & \multirow{2}{*}{$[-1,1]$} & \multirow{2}{*}{$[-1,1]$} & \multirow{2}{*}{$[-1,1]$} & \multirow{2}{*}{$[-1,1]$} & \multirow{2}{*}{$[-1,1]$} \\
    $-\E[Y_i|X_{1i}=0,X_{2i}=1,X_{3i}=1]$ &  &  \\[10pt]
    $\E[Y_i|X_{1i}=0,X_{2i}=0,X_{3i}=0]$ & $[0,1]$ & $[0,1]$ & $[0,1]$ & $[0,1]$ & $[0,0.217]$ & $[0,0.223]$ \\[10pt]
    $\E[Y_i|X_{1i}=1,X_{2i}=0,X_{3i}=0]$ & $[0.116,1]$ & $[0.109,1]$ & $[0.330,1]$ & $[0.250,1]$ & $[0.633,0.992]$ & $[0.633,0.992]$ \\[10pt]
    $\E[Y_i|X_{1i}=0,X_{2i}=1,X_{3i}=0]$ & $[0,1]$ & $[0,1]$ & $[0,1]$ & $[0,1]$ & $[0,1]$ & $[0,1]$ \\[10pt]
    $\E[Y_i|X_{1i}=0,X_{2i}=0,X_{3i}=1]$ & $[0,1]$ & $[0,1]$ & $[0,1]$ & $[0,1]$ & $[0,1]$ & $[0,1]$ \\[10pt]
    $\E[Y_i|X_{1i}=1,X_{2i}=1,X_{3i}=0]$ & $[0,0.940]$ & $[0,0.955]$ & $[0,1]$ & $[0,1]$ & $[0,1]$ & $[0,1]$ \\[10pt]
    $\E[Y_i|X_{1i}=1,X_{2i}=0,X_{3i}=1]$ & $[0,1]$ & $[0,1]$ & $[0,1]$ & $[0,1]$ & $[0,1]$ & $[0,1]$ \\[10pt]
    $\E[Y_i|X_{1i}=0,X_{2i}=1,X_{3i}=1]$ & $[0,1]$ & $[0,1]$ & $[0,1]$ & $[0,1]$ & $[0,1]$ & $[0,1]$ \\[10pt]
    $\E[Y_i|X_{1i}=1,X_{2i}=1,X_{3i}=1]$ & $[0,1]$ & $[0,1]$ & $[0,1]$ & $[0,1]$ & $[0,1]$ & $[0,1]$  \\
    \hline\hline\\[-10pt]
    \end{tabular}}
    \begin{minipage}{\textwidth}
    \fontsize{9}{2}\linespread{1}\selectfont
    \footnotesize{
    \textit{Notes}: Identified sets are population identified sets $D=[L,U]$ under no additional restrictions following the program of Proposition \ref{prop:program_d}. CM sets are bounds obtained following the approach described in Remark \ref{rem:cm}, in which sharp bounds on $\E[Y_i|X_i=x_k,G_i=g]$ are used together with the identified set $P$ for the distribution of $X_i|G_i$ following \eqref{eq:lie_y_x_unknown}. All numbers are rounded to the nearest thousandth. 
    }
    \end{minipage}
\end{table}

\section{Empirical Illustration}
\label{sec5}

One setting in which publicly available data are in aggregate form is standardized exam data. In this section I apply the methodology developed in the previous sections to construct bounds on conditional exam pass rates and conditional white/non-white exam pass rate gaps. I also impose the three additional restrictions discussed in Remark \ref{rem:restrictions}. I find, as suggested by \cite{cho2008cross}, that without any additional assumptions, aggregate data does not have much identifying power for the individual-level parameter. Imposing monotonicity shape restrictions provides some identifying power, although bounds on pass rate gaps are still wide. Using additionally available pass rates by subgroup helps to narrow bounds more than imposing monotonicity shape restrictions. Adding monotonicity shape restrictions to the pass rates by subgroup further tightens bounds. A homogeneity restriction on conditional exam pass rates in different counties results in empty identified sets, suggesting that homogeneity of pass rates across counties is inconsistent with the observed aggregate data.

In this application I focus on exam pass rates for English and math Rhode Island Comprehensive Assessment System (RICAS) exams and student demographic information for the state of Rhode Island in the spring of 2019 over all students in grades 3-8. Data are obtained from the state of Rhode Island Department of Education's Public Assessment Data Portal, available at \url{https://www3.ride.ri.gov/ADP}. Although data is available at the school district level, for computational simplicity I aggregate the data up to the county level. There are 5 counties in Rhode Island, and correspondingly 5 groups in the data. 

I estimate identified sets for pass rates and white/non-white pass rate gaps conditional on three covariates: race (indicator $white_i$ for being white, where non-white students are Hispanic, Black, Asian, Pacific Islander, Native American, or two or more races), economically disadvantaged status (indicator $econ_i$ for being economically disadvantaged, as defined by the Rhode Island Department of Education), and English-language learner (ELL) status (indicator $ELL_i$ for students who are currently English-language learners or were English-language learners in the past 3 years). 

In addition to the aggregate data of Assumption \ref{ass:1}.iv, the Rhode Island Department of Education also makes available pass rates conditional on each of the three covariates for school districts with a sufficiently high number of students in each subgroup. I use the following three subgroup pass rates: $\E[pass_i|white_i=1,G_i=g], \E[pass_i|econ_i=0,G_i=g]$, and $\E[pass_i|ELL_i=0,G_i=g]$. I drop all districts that are missing any of these subgroup pass rates before aggregating to the county level. The subgroup pass rates imply additional restrictions as discussed in Remark \ref{rem:restrictions}.

I first estimate bounds without any further assumptions, then impose additional monotonicity shape restrictions on the conditional pass rate.
Motivated by test score gaps that have been documented between rich and poor students \citep{NYT2012, NYT2015}, I consider the monotonicity restriction that for each value of $(white_i, ELL_i)$ and each county $g$ that the average pass rate is lower for economically disadvantaged students:
\begin{equation} \label{eq:mono1}
    \E[pass_i|econ_i=1, white_i, ELL_i, G_i = g] - \E[pass_i|econ_i=0, white_i, ELL_i, G_i = g] \leq 0.
\end{equation}
For English exams, I impose an additional monotonicity restriction that for each value of $(white_i, econ_i)$ and each county $g$ the average pass rate is lower for English-language learner students than for non-English-language learner students:
\begin{equation} \label{eq:mono2}
    \E[pass_i|ELL_i=1, white_i, econ_i, G_i = g] - \E[pass_i|ELL_i=0, white_i, econ_i, G_i = g] \leq 0.
\end{equation}
I then estimate bounds using the additional subgroup pass rates, without imposing monotonicity. Finally, I estimate bounds using the additional subgroup pass rates under monotonicity.

I first present estimated bounds on math exam white/non-white pass rate gaps in Table \ref{tab:dat_math_bds} and on English exam white/non-white pass rate gaps in Table \ref{tab:dat_ela_bds}. 90\% confidence intervals constructed as in Section \ref{sec3_2} are displayed below the bound estimates in curly brackets. For both types of exam, bounds on the white/non-white pass rate gaps reported in Column 1 are wide and either uninformative or close to uninformative. Imposing the additional monotonicity restrictions helps narrow the bounds for some parameters, reported in Column 2, but bounds are still wide. Using additional subgroup pass rates without monotonicity in Column 3 further narrows the bounds for those same parameters. Adding monotonicity to additional subgroup pass rates in Column 4 produces the tightest bounds, especially for white/non-white pass rate gaps among students who are economically disadvantaged but not English-language learners. For example, among economically disadvantaged students who are not English-language learners, the white/non-white English exam pass rate gap is estimated to be no lower than -30\% and no higher than 64\%. Bounds on pass rate gaps among English-language learners are never informative, as expected due to the small percentage of English-language learners in Rhode Island.

\begin{table}[!htbp] 
    \caption{Bounds on Rhode Island conditional white/non-white math exam pass rate differences}
    \label{tab:dat_math_bds}
    \centering
    \resizebox{\textwidth}{!}{
    \begin{tabular}{ccccc}
    \hline\hline 
    & & \\[-12pt]
     & No restrictions & Monotonicity & Subgroup & Subgroup and \\
    & & & & monotonicity \\
    Parameter & (1) & (2) & (3) & (4) \\
    \hline
    $\E \big [pass_{i} \big | white_{i} = 1, econ_{i} = 0, ELL_{i} = 0\big ]$ & $[-0.974, 0.897]$ & $[-0.898, 0.897]$ & $[-0.695, 0.734]$ & $[-0.623, 0.523]$ \\
    $-\E \big [pass_{i} \big | white_{i} = 0, econ_{i} = 0, ELL_{i} = 0\big ]$ & $\{-0.993, 0.940\}$ & $\{-0.924, 0.940\}$ & $\{-0.812, 0.817\}$ & $\{-0.689, 0.792\}$ \\[10pt]
    $\E \big [pass_{i} \big | white_{i} = 1, econ_{i} = 1$, $ELL_{i} = 0\big ]$ & $[-1, 1]$  & $[-0.990, 0.925]$ & $[-0.625, 0.738]$ & $[-0.260, 0.522]$ \\
    $-\E \big [pass_{i} \big | white_{i} = 0, econ_{i} = 1$, $ELL_{i} = 0\big ]$ & $\{-1, 1\}$ & $\{-1, 0.954\}$ & $\{-0.761, 0.896\}$ & $\{-0.381, 0.638\}$ \\[10pt]
    $\E \big [pass_{i} \big | white_{i} = 1, econ_{i} = 0$, $ELL_{i} = 1\big ]$ & $[-1,1]$ & $[-1,1]$ & $[-1, 1]$ & $[-1, 1]$ \\
    $-\E \big [pass_{i} \big | white_{i} = 0, econ_{i} = 0$, $ELL_{i} = 1\big ]$ & $\{-1, 1\}$ & $\{-1, 1\}$ & $\{-1, 1\}$ & $\{-1, 1\}$ \\[10pt]
    $\E \big [pass_{i} \big | white_{i} = 1, econ_{i} = 1$, $ELL_{i} = 1\big ]$ & $[-1, 1]$  & $[-1, 1]$ & $[-1, 1]$ & $[-1, 1]$ \\
    $-\E \big [pass_{i} \big | white_{i} = 0, econ_{i} = 1$, $ELL_{i} = 1\big ]$ & $\{-1, 1\}$ & $\{-1, 1\}$ & $\{-1, 1\}$ & $\{-1, 1\}$ \\
    \hline\hline\\[-10pt]
    \end{tabular}}
    \begin{minipage}{\textwidth}
    \fontsize{9}{2}\linespread{1}\selectfont
    \footnotesize{
    \textit{Notes}: Reported intervals are bound estimates, that is, sample analogs of the identified set $D = [L,U]$. 90\% confidence intervals are constructed following the inference procedure of Section \ref{sec3_2} and are below the estimated intervals in curly brackets. Bounds with no restrictions follow the program of Proposition \ref{prop:program_d}. Monotonicity bounds impose additional monotonicity restriction \eqref{eq:mono1}. Subgroup bounds use additional data on pass rates among subgroups of students to impose additional restrictions, as described in Remark \ref{rem:restrictions} and Section \ref{sec5}. Subgroup and monotonicity bounds impose all restrictions implied by subgroup pass rates and monotonicity restriction \eqref{eq:mono1}. All numbers are rounded to the nearest thousandth.
    }
    \end{minipage}
\end{table}

\begin{table}[!htbp] 
    \caption{Bounds on Rhode Island conditional white/non-white English exam pass rate differences}
    \label{tab:dat_ela_bds}
    \centering
    \resizebox{\textwidth}{!}{
    \begin{tabular}{ccccc}
    \hline\hline 
    & & \\[-12pt]
     & No restrictions & Monotonicity & Subgroup & Subgroup and \\
    & & & & monotonicity \\
    Parameter & (1) & (2) & (3) & (4) \\
    \hline
    $\E \big [pass_{i} \big | white_{i} = 1, econ_{i} = 0, ELL_{i} = 0\big ]$ & $[-0.940, 0.987]$ & $[-0.830, 0.98 7]$ & $[-0.607, 0.820]$ & $[-0.507, 0.513]$ \\
    $-\E \big [pass_{i} \big | white_{i} = 0, econ_{i} = 0, ELL_{i} = 0\big ]$ & $\{-0.959, 1\}$ & $\{-0.853, 1\}$ & $\{-0.729, 0.902\}$ & $\{-0.549, 0.815\}$ \\[10pt]
    $\E \big [pass_{i} \big | white_{i} = 1, econ_{i} = 1$, $ELL_{i} = 0\big ]$ & $[-1, 1]$  & $[-0.983,0.952]$ & $[-0.626, 0.816]$ & $[-0.294, 0.635]$ \\
    $-\E \big [pass_{i} \big | white_{i} = 0, econ_{i} = 1$, $ELL_{i} = 0\big ]$ & $\{-1, 1\}$ & $\{-1, 0.973\}$ & $\{-0.757, 0.945\}$ & $\{-0.443, 0.763\}$ \\[10pt]
    $\E \big [pass_{i} \big | white_{i} = 1, econ_{i} = 0$, $ELL_{i} = 1\big ]$ & $[-1,1]$ & $[-1,1]$ & $[-1, 1]$ & $[-1, 1]$ \\
    $-\E \big [pass_{i} \big | white_{i} = 0, econ_{i} = 0$, $ELL_{i} = 1\big ]$ & $\{-1, 1\}$ & $\{-1, 1\}$ & $\{-1, 1\}$ & $\{-1, 1\}$ \\[10pt]
    $\E \big [pass_{i} \big | white_{i} = 1, econ_{i} = 1$, $ELL_{i} = 1\big ]$ & $[-1, 1]$  & $[-1, 1]$ & $[-1, 1]$ & $[-1, 1]$ \\
    $-\E \big [pass_{i} \big | white_{i} = 0, econ_{i} = 1$, $ELL_{i} = 1\big ]$ & $\{-1, 1\}$ & $\{-1, 1\}$ & $\{-1, 1\}$ & $\{-1, 1\}$ \\
    \hline\hline\\[-10pt]
    \end{tabular}}
    \begin{minipage}{\textwidth}
    \fontsize{9}{2}\linespread{1}\selectfont
    \footnotesize{\textit{Notes}: Reported intervals are bound estimates, that is, sample analogs of the identified set $D = [L,U]$. 90\% confidence intervals are constructed following the inference procedure of Section \ref{sec3_2} and are below the estimated intervals in curly brackets. Bounds with no restrictions follow the program of Proposition \ref{prop:program_d}. Monotonicity bounds impose additional monotonicity restrictions \eqref{eq:mono1} and \eqref{eq:mono2}. Subgroup bounds use additional data on pass rates among subgroups of students to impose additional restrictions, as described in Remark \ref{rem:restrictions} and Section \ref{sec5}. Subgroup and monotonicity bounds impose all restrictions implied by subgroup pass rates and monotonicity restrictions \eqref{eq:mono1} and \eqref{eq:mono2}. All numbers are rounded to the nearest thousandth.
    }
    \end{minipage}
\end{table}

In Table \ref{tab:dat_math_bds_indv} I present estimated bounds on math exam conditional pass rates and in Table \ref{tab:dat_ela_bds_indv} I present estimated bounds on English exam conditional pass rates. 90\% confidence intervals constructed as in Section \ref{sec3_2} are displayed below the bound estimates in curly brackets. Just as for the white/non-white pass rate gaps, the additional restrictions help to narrow the bounds on pass rates for students who are not English-language learners. For example, the math exam pass rate among non-white students who are economically disadvantaged and not English-language learners is estimated to be bounded above by 29\%, while the math exam pass rate among white students who are not economically disadvantaged and not English-language learners is estimated to be between 37\% and 72\%. Similarly, the English exam pass rate among non-white students who are economically disadvantaged and not English-language learners is estimated to be bounded above by 36\%, while the math exam pass rate among white students who are not economically disadvantaged and not English-language learners is estimated to be between 50\% and 80\%.

\begin{table}[!htbp] 
    \caption{Bounds on Rhode Island math exam conditional pass rates}
    \label{tab:dat_math_bds_indv}
    \centering
    \resizebox{\textwidth}{!}{
    \begin{tabular}{ccccc}
    \hline\hline 
    & & \\[-12pt]
     & No restrictions & Monotonicity & Subgroup & Subgroup and \\
    & & & & monotonicity \\
    Parameter & (1) & (2) & (3) & (4) \\
    \hline
    \multirow{2}{*}{$\E \big [pass_{i} \big | white_{i} = 0, econ_{i} = 0, ELL_{i} = 0\big ]$} & $[0,1]$ & $[0,1]$ & $[0.005,1]$ & $[0.151,1]$\\
    & $\{0,1\}$ & $\{0,1\}$ & $\{0,1\}$ & $\{0,1\}$ \\[10pt]
    \multirow{2}{*}{$\E \big [pass_{i} \big | white_{i} = 1, econ_{i} = 0, ELL_{i} = 0\big ]$} & $[0.026,0.897]$ & $[0.102,0.897]$ & $[0.309,0.738]$ & $[0.377,0.720]$\\
    & $\{0.007,0.940\}$ & $\{0.076,0.940\}$ & $\{0.188,0.821\}$ & $\{0.310, 0.858\}$ \\[10pt]
    \multirow{2}{*}{$\E \big [pass_{i} \big | white_{i} = 0, econ_{i} = 1$, $ELL_{i} = 0\big ]$} & $[0,1]$ & $[0,0.990]$ & $[0,0.626]$ & $[0,0.287]$\\
    & $\{0,1\}$ & $\{0,1\}$ & $\{0,0.761\}$ & $\{0,0.540\}$ \\[10pt]
    \multirow{2}{*}{$\E \big [pass_{i} \big | white_{i} = 0, econ_{i} = 0$, $ELL_{i} = 1\big ]$} & $[0,1]$ & $[0,1]$ & $[0,1]$ & $[0,1]$\\
    & $\{0,1\}$ & $\{0,1\}$ & $\{0,1\}$ & $\{0,1\}$ \\[10pt]
    \multirow{2}{*}{$\E \big [pass_{i} \big | white_{i} = 1, econ_{i} = 1$, $ELL_{i} = 0\big ]$} & $[0,1]$ & $[0,0.925]$ & $[0.001,0.738]$ & $[0.035,0.523]$\\
    & $\{0,1\}$ & $\{0,0.954\}$ & $\{0,0.896\}$ & $\{0,0.647\}$ \\[10pt]
    \multirow{2}{*}{$\E \big [pass_{i} \big | white_{i} = 1, econ_{i} = 0$, $ELL_{i} = 1\big ]$} & $[0,1]$ & $[0,1]$ & $[0,1]$ & $[0,1]$\\
    & $\{0,1\}$ & $\{0,1\}$ & $\{0,1\}$ & $\{0,1\}$ \\[10pt]
    \multirow{2}{*}{$\E \big [pass_{i} \big | white_{i} = 0, econ_{i} = 1$, $ELL_{i} = 1\big ]$} & $[0,1]$ & $[0,1]$ & $[0,1]$ & $[0,1]$\\
    & $\{0,1\}$ & $\{0,1\}$ & $\{0,1\}$ & $\{0,1\}$ \\[10pt]
    \multirow{2}{*}{$\E \big [pass_{i} \big | white_{i} = 1, econ_{i} = 1$, $ELL_{i} = 1\big ]$} & $[0,1]$ & $[0,1]$ & $[0,1]$ & $[0,1]$\\
    & $\{0,1\}$ & $\{0,1\}$ & $\{0,1\}$ & $\{0,1\}$ \\
    \hline\hline\\[-10pt]
    \end{tabular}}
    \begin{minipage}{\textwidth}
    \fontsize{9}{2}\linespread{1}\selectfont
    \footnotesize{
    \textit{Notes}: Reported intervals are bound estimates, that is, sample analogs of the identified set $D = [L,U]$. 90\% confidence intervals are constructed following the inference procedure of Section \ref{sec3_2} and are below the estimated intervals in curly brackets. Bounds with no restrictions follow the program of Proposition \ref{prop:program_d}. Monotonicity bounds impose additional monotonicity restriction \eqref{eq:mono1}. Subgroup bounds use additional data on pass rates among subgroups of students to impose additional restrictions, as described in Remark \ref{rem:restrictions} and Section \ref{sec5}. Subgroup and monotonicity bounds impose all restrictions implied by subgroup pass rates and monotonicity restriction \eqref{eq:mono1}. All numbers are rounded to the nearest thousandth.}
    \end{minipage}
    \end{table}

\begin{table}[!htbp] 
    \caption{Bounds on Rhode Island English exam conditional pass rates}
    \label{tab:dat_ela_bds_indv}
    \centering
    \resizebox{\textwidth}{!}{
    \begin{tabular}{ccccc}
    \hline\hline 
    & & \\[-12pt]
     & No restrictions & Monotonicity & Subgroup & Subgroup and \\
    & & & & monotonicity \\
    Parameter & (1) & (2) & (3) & (4) \\
    \hline
    \multirow{2}{*}{$\E \big [pass_{i} \big | white_{i} = 0, econ_{i} = 0, ELL_{i} = 0\big ]$} & $[0,1]$ & $[0,1]$ & $[0.003,1]$ & $[0.234,1]$\\
    & $\{0,1\}$ & $\{0,1\}$ & $\{0,1\}$ & $\{0.022,1\}$ \\[10pt]
    \multirow{2}{*}{$\E \big [pass_{i} \big | white_{i} = 1, econ_{i} = 0, ELL_{i} = 0\big ]$} & $[0.060,0.987]$ & $[0.170,0.987]$ & $[0.572,0.823]$ & $[0.500,0.795]$\\
    & $\{0.041,1\}$ & $\{0.147,1\}$ & $\{0.271,0.906\}$ & $\{0.450, 0.935\}$ \\[10pt]
    \multirow{2}{*}{$\E \big [pass_{i} \big | white_{i} = 0, econ_{i} = 1$, $ELL_{i} = 0\big ]$} & $[0,1]$ & $[0,0.983]$ & $[0,0.643]$ & $[0,0.358]$\\
    & $\{0,1\}$ & $\{0,1\}$ & $\{0,0.757\}$ & $\{0,0.577\}$ \\[10pt]
    \multirow{2}{*}{$\E \big [pass_{i} \big | white_{i} = 0, econ_{i} = 0$, $ELL_{i} = 1\big ]$} & $[0,1]$ & $[0,1]$ & $[0,1]$ & $[0,1]$\\
    & $\{0,1\}$ & $\{0,1\}$ & $\{0,1\}$ & $\{0,1\}$ \\[10pt]
    \multirow{2}{*}{$\E \big [pass_{i} \big | white_{i} = 1, econ_{i} = 1$, $ELL_{i} = 0\big ]$} & $[0,1]$ & $[0,0.952]$ & $[0.019,0.816]$ & $[0.196,0.638]$\\
    & $\{0,1\}$ & $\{0,0.973\}$ & $\{0,0.945\}$ & $\{0,0.777\}$ \\[10pt]
    \multirow{2}{*}{$\E \big [pass_{i} \big | white_{i} = 1, econ_{i} = 0$, $ELL_{i} = 1\big ]$} & $[0,1]$ & $[0,1]$ & $[0,1]$ & $[0,1]$\\
    & $\{0,1\}$ & $\{0,1\}$ & $\{0,1\}$ & $\{0,1\}$ \\[10pt]
    \multirow{2}{*}{$\E \big [pass_{i} \big | white_{i} = 0, econ_{i} = 1$, $ELL_{i} = 1\big ]$} & $[0,1]$ & $[0,1]$ & $[0,1]$ & $[0,1]$\\
    & $\{0,1\}$ & $\{0,1\}$ & $\{0,1\}$ & $\{0,1\}$ \\[10pt]
    \multirow{2}{*}{$\E \big [pass_{i} \big | white_{i} = 1, econ_{i} = 1$, $ELL_{i} = 1\big ]$} & $[0,1]$ & $[0,1]$ & $[0,1]$ & $[0,1]$\\
    & $\{0,1\}$ & $\{0,1\}$ & $\{0,1\}$ & $\{0,1\}$ \\
    \hline\hline\\[-10pt]
    \end{tabular}}
    \begin{minipage}{\textwidth}
    \fontsize{9}{2}\linespread{1}\selectfont
    \footnotesize{
    \textit{Notes}: Reported intervals are bound estimates, that is, sample analogs of the identified set $D = [L,U]$. 90\% confidence intervals are constructed following the inference procedure of Section \ref{sec3_2} and are below the estimated intervals in curly brackets. Bounds with no restrictions follow the program of Proposition \ref{prop:program_d}. Monotonicity bounds impose additional monotonicity restrictions \eqref{eq:mono1} and \eqref{eq:mono2}. Subgroup bounds use additional data on pass rates among subgroups of students to impose additional restrictions, as described in Remark \ref{rem:restrictions} and Section \ref{sec5}. Subgroup and monotonicity bounds impose all restrictions implied by subgroup pass rates and monotonicity restrictions \eqref{eq:mono1} and \eqref{eq:mono2}. All numbers are rounded to the nearest thousandth.}
    \end{minipage}
\end{table}

While each of these bounds are sharp for each of the corresponding parameters, as proved in Proposition \ref{prop:program_d}, note that the bounds are not jointly sharp across all parameters in the tables. In fact, the Minkowski difference of bounds on conditional pass rates in Tables \ref{tab:dat_math_bds_indv} and \ref{tab:dat_ela_bds_indv} are sometimes wider than the bounds on the corresponding pass rate gaps in Tables \ref{tab:dat_math_bds} and \ref{tab:dat_ela_bds}, especially with additional subgroup pass rates under monotonicity. If the bounds were jointly sharp across all parameters in Tables \ref{tab:dat_math_bds_indv} and \ref{tab:dat_ela_bds_indv} then the Minkowski difference of the bounds on conditional pass rates should be equal to the bounds on the corresponding pass rate gaps. That the Minkowski difference is wider highlights an advantage of my method, which is that my method directly produces sharp bounds for the linear combination of conditional pass rates.

Another restriction that I could impose, as discussed in Remark \ref{rem:restrictions}, is a homogeneity restriction on the conditional pass rate across counties. This restriction imposes that for every value of $(white_i, econ_i, ELL_i)$, $\E[pass_i|white_i, econ_i, ELL_i, G_i = g] = \E[pass_i|white_i, econ_i, ELL_i, G_i = g']$ for all pairs of counties $g$ and $g'$. This assumption, which is very strong, can be interpreted as imposing that county exam pass rates differ only because of differences in student characteristics and not because of, for example, differences in county value-added to student education.
When I estimate bounds for both math and English exam white/non-white pass rate gaps, I obtain empty sets. This means that homogeneity is inconsistent with the observed data, as discussed in Remark \ref{rem:restrictions}.

\section{Conclusion}
\label{sec6} 

In this paper I consider the problem of identifying linear combinations of conditional mean outcomes when the data that is observed is marginal information on each individual-level variable over many groups, which I call aggregate data. This model of aggregate data necessitates a nontrivial extension of existing methods in the ecological inference literature. I develop a partial identification methodology to construct bounds by solving an optimization problem that considers all joint distributions of individual-level variables that are consistent with the observed marginal information. This approach easily accommodates additional model restrictions like polyhedral shape restrictions. I propose a plug-in estimator of the identified set that is consistent and a valid method to construct confidence intervals that uses aggregate data only. I demonstrate the performance of the method in simulation studies and apply the method to Rhode Island standardized exam data.

\section*{Acknowledgements}

I am especially grateful to Edward Vytlacil, Isaiah Andrews, and Anna Mikusheva for their guidance and advice. I thank Editor Yuya Sasaki, the Associate Editor, two anonymous referees, Alberto Abadie, Hyungsik Roger Moon, Whitney Newey, Jesse Shapiro, Elie Tamer, Vod Vilfort, and participants in the MIT econometrics lunch seminar for their helpful comments and suggestions. All errors are my own.

\section*{Funding}

This paper is based upon work supported by the National Science Foundation Graduate Research Fellowship Program under Grant No. 2141064.

\section*{Disclosure Statement}

There are no competing interests to declare.

%\newpage

% references
\singlespacing
{
\bibliographystyle{chicago}
\bibliography{ref}
}

\clearpage

\newpage

% Appendix

\appendix
\onehalfspacing
% Restart Footnotes count
\newcommand{\AppendixPrefix}{A}
% Restart Figures, tables and equations and label with appendix section number
\renewcommand{\thefigure}{\thesection.\arabic{figure}}
\counterwithin{figure}{section}
\renewcommand{\thetable}{\thesection.\arabic{table}}
\counterwithin{table}{section}
\renewcommand{\theequation}{\thesection.\arabic{equation}}
\counterwithin{equation}{section}
\renewcommand{\thecorollary}{\thesection.\arabic{corollary}}
\counterwithin{corollary}{section}
\renewcommand{\theproposition}{\thesection.\arabic{proposition}}
\counterwithin{proposition}{section}

\section{Appendix: Additional Results}
\label{app:results}

\subsection{Bounds using the \cite{cross2002regressions} method}
\label{app:cm}

\cite{cross2002regressions} and \cite{cho2008cross} provide sharp bounds on $\E[Y_i|X_i, G_i]$ jointly over the support of $X_i$ when one knows the probability distribution of $Y_i|G_i$ and the probability distribution of $X_i|G_i$. In my setting instead of the distribution of $Y_i|G_i$ I observe $\E[Y_i|G_i]$ and instead of the joint distribution of $X_i|G_i$ I observe the marginal distribution of each $X_{\ell i}|G_i$. 

However, note that observing $\E[Y_i|G_i]$ for a random variable supported on $[y_\ell, y_u]$ is equivalent to observing the distribution of $\tilde Y_i|G_i$, for $\tilde Y_i$ a random variable that takes value $y_\ell$ given $G_i = g$ with probability $\frac{\E[Y_i|G_i=g] - y_u}{y_\ell - y_u}$ and value $y_u$ with probability $1- \frac{\E[Y_i|G_i=g] - y_u}{y_\ell - y_u}$. In other words, $Y_i$ and $\tilde Y_i$ are indistinguishable given the aggregate data and assumptions.

In addition, it can be shown analogously to the proof of Proposition \ref{prop:program_d} that set $P$ defined by \eqref{eq:ident_x} is the sharp identified set for the joint distribution of $X_i|G_i$ given the marginal distribution of each $X_{\ell i}|G_i$.
So the union of the bounds obtained from the \cite{cross2002regressions} method with the distribution $\tilde Y_i|G_i$ and each candidate distribution for $X_i|G_i$ in the set $P$ is a sharp identified set for the parameter $\E[Y_i|X_i=x_k,G_i=g]$ jointly over the support of $X_i$.

\section{Appendix: Proofs}
\label{app:proofs}

\subsection{Proof of Proposition \ref{prop:program_d}} \label{app_proof_prop:program_d}

\begin{proof}
    I first show that $D = [L,U]$.

    The optimization problem to obtain $P$ is minimized at $(p_{11},\dots,p_{KG}) = (\P[X_i=x_1|G_i=1], \dots, \P[X_i=x_K|G_i=G])$ because all constraints are satisfied with the last constraint satisfied for $v_{rg}^+ = v_{rg}^- = 0$ for all $r$ and $g$. Any other minimizer thus satisfies all constraints with the last one holding for $v_{rg}^+ = v_{rg}^- = 0$. These are exactly the elements that satisfy the constraints in \eqref{eq:ident_x}.

    It is clear from the definition of $D$ in \eqref{eq:ident_d} that $L = \inf D$ and $U = \sup D$.
    That $D = [L,U]$ follows if I show $D$ is an interval and if I show $D$ attains its infimum and supremum. 
    
    To show $D$ is an interval, I will show that for arbitrary $\delta \in D$, any value between $\delta$ and $\sum_{k=1}^K \lambda_k \E[Y_i|G_i = g]$ is contained in $D$. In particular, $\delta$ implies corresponding $(d_1,\dots,d_K)$, $(p_{11},\dots,p_{KG})$, and $(c_{11},\dots,c_{KG})$ that satisfy the conditions of \eqref{eq:ident_d}. Because $\sum_{k=1}^K p_{kg} = 1$ for all $g$, $\E[Y_i|G_i=g] = \sum_{k=1}^K \E[Y_i|G_i=g] p_{kg}$ also holds for all $g$. 

    Thus for any $t \in [0,1]$ I can construct 
    \begin{align*}
        \hat c_{kg} &= tc_{kg} + (1-t) \E[Y_i|G_i=g] \in [y_\ell, y_u].
    \end{align*}
    If $\sum_{g=1}^G \P[G_i = g] p_{kg} > 0$ for a given $k$, let $\hat d_k = \sum_{g=1}^G \frac{\P[G_i = g]p_{kg}\hat c_{kg}}{\sum_{g=1}^G \P[G_i = g]p_{kg}}$, which is contained in $[y_\ell, y_u]$. If instead $\sum_{g=1}^G \P[G_i = g] p_{kg} = 0$ for a given $k$, let $\hat d_k = t\delta + (1-t) \E[Y_i|G_i = g]$. Then the value
    $\sum_{k=1}^K \lambda_k \hat d_k = t \delta + (1-t) \sum_{k=1}^K \lambda_k \E[Y_i|G_i = g]$ is contained in $D$ for any $t \in [0,1]$. 

    To show $D$ attains its infimum and supremum, following the proof of Proposition \ref{prop:consistency} I can conclude that the set of $\{p_{kg}\}$, $\{c_{kg}\}$, and $\{d_k\}$ that satisfy the constraints of sets $P$ and $D$ are compact-valued for any given data consistent with Assumption \ref{ass:1}. Because $\sum_{k=1}^K \lambda_k d_k$ is a continuous function of $\{d_k\}$ over the compact set, the infimum and supremum of $D$ are attained in $D$.

    I now show that $D$ is the sharp identified set. To do so I show that every value in $D$ is attained by some distribution for the data $(Y_i, X_i, G_i)$ that is consistent with Assumption \ref{ass:1} and the observed data. Let $\delta \in D$ be arbitrary. By definition there exist distributions $\{p_{kg}\}$ for $X_i|G_i$ and values $\{c_{kg}\}$ for $\E[Y_i|X_i=x_k,G_i=g]$ and $\{d_k\}$ for $\E[Y_i|X_i = x_k]$ that are consistent with the observed aggregate data and Assumption \ref{ass:1}. Because I observe the distribution of $G_i$, this together with $\{p_{kg}\}$ pins down a distribution for $(X_i, G_i)$. For each $k$ and $g$ such that $p_{kg} > 0$ I can take any distribution for $Y_i|X_i=x_k, G_i=g$ that is consistent with the moments $c_{kg}$ and $d_k$ and the bounded support assumption $Y_i \in [y_\ell, y_u]$. This will pin down a joint distribution for $(Y_i, X_i, G_i)$ because for any $k,g$ such that $p_{kg}=0$, I already know $\P[Y_i = y, X_i=x_k, G_i=g] = 0$ for any $y \in [y_\ell, y_u]$. 
    Under this distribution, I know by definition that $\sum_{k=1}^K \lambda_k \E[Y_i|X_i=x_k] = \delta$.
    Thus $D$ is sharp.
\end{proof}

\subsection{Proof of Proposition \ref{prop:consistency}} \label{app_proof_prop:consistency}

\begin{proof}
Note that the objective functions of programs \eqref{eq:opt_lower} and \eqref{eq:opt_upper} in Proposition \ref{prop:program_d} are continuous in $\{d_k\}$. Thus from Proposition \ref{prop:program_d}, convergence of the bounds follows from Berge's maximum theorem if the correspondence
\begin{multline*}
    \Gamma\left(\{\alpha_g\}, \{\beta_{k, \ell, g}\} \right) = \bigg\{ \{d_k\}: d_k \in [y_\ell, y_u] ~ \forall k \text{ and } \exists (\{p_{kg}\}, \{c_{kg}\}) \in C\left(\{\alpha_g\}, \{\beta_{k, \ell, g}\} \right) \; \\
    \text{s.t. } d_k \sum_{g=1}^G \P[G_i=g] p_{kg} = \sum_{g=1}^G \P[G_i=g] p_{kg}c_{kg} ~ \forall k \bigg \}
\end{multline*}
is continuous and compact-valued at any data $\left(\{\alpha_g\}, \{\beta_{k, \ell, g}\} \right)$ such that 
\begin{enumerate}[label=(\roman*)]
    \item each $\alpha_g \in [y_\ell, y_u]$,
    \item $\{\beta_{k, \ell, g}\}$ are valid marginal distributions of $X_i|G_i=g$,
    \item the correspondence $\Gamma \left(\{\alpha_g\}, \{\beta_{k, \ell, g}\} \right)$ is nonempty,
\end{enumerate}
where
\begin{multline*}
    C\left(\{\alpha_g\}, \{\beta_{k, \ell, g}\} \right) = \bigg\{ \{p_{kg}\}, \{c_{kg}\}: \; p_{kg} \geq 0 ~\forall k,g, \; c_{kg} \in [y_\ell, y_u] ~ \forall k,g, \; \sum_{k=1}^K p_{kg} = 1 ~ \forall g, \\
    \beta_{k,\ell,g} = \sum_{j=1}^K \mathbbm{1}\{x_{j,\ell} = x_{k,\ell}\} p_{jg} ~ \forall \ell,k,g, \text{ and } \alpha_g = \sum_{k=1}^K c_{kg}p_{kg} ~ \forall g \bigg \}.
\end{multline*}
If $\Gamma$ is continuous and compact-valued, the Theorem of the Maximum implies that both the lower and the upper bound are continuous functions of the data. The data are sample means, which converge in probability by the law of large numbers to their respective population parameters under Assumption \ref{ass:5}.
Then the result follows from the continuous mapping theorem.

Note first that when $\Gamma$ is nonempty, by the same proof in the proof of Proposition \ref{prop:program_d} that the identified set $D$ is a closed interval, $\Gamma$ must be compact-valued. To show $\Gamma$ is continuous, I must show it is both upper and lower hemicontinuous.

To show upper hemicontinuity, I will consider the correspondence
\begin{multline*}
    \Psi\left(\{\alpha_g\}, \{\beta_{k, \ell, g}\} \right) = \bigg\{ \{p_{kg}\}, \{c_{kg}\}, \{d_k\}: d_k \in [y_\ell, y_u] ~ \forall k, \; p_{kg} \geq 0 ~\forall k,g, \; c_{kg} \in [y_\ell, y_u] ~ \forall k,g, \\
    \sum_{k=1}^K p_{kg} = 1 ~ \forall g, \; \beta_{k,\ell,g} = \sum_{j=1}^K \mathbbm{1}\{x_{j,\ell} = x_{k,\ell}\} p_{jg} ~ \forall \ell,k,g, \; \alpha_g = \sum_{k=1}^K c_{kg}p_{kg} ~ \forall g,  \\
    \text{and } d_k \sum_{g=1}^G \P[G_i=g] p_{kg} = \sum_{g=1}^G \P[G_i=g] p_{kg}c_{kg} ~ \forall k \bigg \}.
\end{multline*}
Note that the projection of $\Psi$ onto the coordinates corresponding to $\{d_k\}$ is exactly $\Gamma$. By the open set definition of upper hemicontinuity, if I can show $\Psi$ is upper hemicontinuous then $\Gamma$ is also upper hemicontinuous because projection is continuous.

To show upper hemicontinuity of $\Psi$, consider an arbitrary sequence of data $\left(\{\alpha_g^n\}, \{\beta_{k, \ell, g}^n\} \right)$ that converges to $\left(\{\alpha_g\}, \{\beta_{k, \ell, g}\} \right)$ as $n \to \infty$ and satisfies the conditions above for each $n$. Form another arbitrary sequence $\left( \{p_{kg}^n\}, \{c_{kg}^n\} , \{d_k^n\}\right) \to \left( \{p_{kg}\}, \{c_{kg}\} , \{d_k\}\right)$ such that $\left( \{p_{kg}^n\}, \{c_{kg}^n\} , \{d_k^n\}\right) \in \Psi \left(\{\alpha_g^n\}, \{\beta_{k, \ell, g}^n\} \right)$ for each $n$. I'd like to show that $\left( \{p_{kg}\}, \{c_{kg}\} , \{d_k\}\right) \in \Psi \left(\{\alpha_g\}, \{\beta_{k, \ell, g}\} \right)$.
This follows immediately when looking at the constraints of correspondence $\Psi$ from the following properties of sequences: $a_n \leq b_n \Rightarrow \lim_n a_n \leq \lim_n b_n$, $\lim_n (a_n+b_n) = \lim_n a_n + \lim_n b_n$, and $\lim_n a_nb_n = \left(\lim_n a_n \right)\left(\lim_n b_n\right)$.

To show lower hemicontinuity, consider an arbitrary sequence of data $\left(\{\alpha_g^n\}, \{\beta_{k, \ell, g}^n\} \right)$ that converges to $\left(\{\alpha_g\}, \{\beta_{k, \ell, g}\} \right)$ as $n \to \infty$ and satisfies the conditions above for each $n$. Let $\{d_k\} \in \Gamma \left(\{\alpha_g\}, \{\beta_{k, \ell, g}\} \right)$ be arbitrary. I'd like to show there exists some sequence $\{d_k^t\} \to \{d_k\}$ as $t \to \infty$ such that $\{d_k^t\} \in \Gamma \left(\{\alpha_g^{n_t}\}, \{\beta_{k, \ell, g}^{n_t}\} \right)$ for $n_t \to \infty$ a subsequence of $\{n\}$.

Note that there are corresponding $\left( \{p_{kg}\}, \{c_{kg}\} \right) \in C \left(\{\alpha_g\}, \{\beta_{k, \ell, g}\} \right)$ such that the constraint $d_k \sum_{g=1}^G \P[G_i=g] p_{kg} = \sum_{g=1}^G \P[G_i=g] p_{kg} c_{kg}$ holds for all $k$. If $\sum_{g=1}^G \P[G_i=g] p_{kg} = 0$ for a given $k$ then I know that each $p_{kg} = 0$ for all $g$ given that value of $k$. Then any $\{c_{kg}\}$ such that $c_{kg} \in [y_\ell, y_u]$ for all $g$ for that given value of $k$ satisfies the constraints of $C$ and that constraint of $\Gamma$ for that $\{p_{kg}\}$. In this case that $\sum_{g=1}^G \P[G_i=g] p_{kg} = 0$ for a given $k$, choose $\{c_{kg}\}$ such that $c_{kg} = d_k$ for all $g$ for that given value of $k$.

The system of equations $\sum_{k=1}^K p_{kg} = 1 ~ \forall g$ and $\beta_{k,\ell,g} = \sum_{j=1}^K \mathbbm{1}\{x_{j,\ell} = x_{k,\ell}\} p_{jg} ~ \forall \ell,k,g$ has strictly less than $K$ unique noncollinear equations if $\{\beta_{k,\ell,g}\}$ are valid marginal distributions of $X_i|G_i = g$. Then by the Moore-Penrose inverse formula for the solution to an underdetermined system of linear equations, there are infinitely many solutions to the system of equations and a solution $\{p_{kg}\}$ to the system of equations is continuous in $\{\beta_{k,\ell,g}\}$. So for each $t$ there exists large enough $n_t$ such that $\{p_{kg}^t\}$ is arbitrarily close to $\{p_{kg}\}$, $\sum_{k=1}^K p_{kg}^t = 1 ~ \forall g$ and $\beta_{k,\ell,g}^{n_t} = \sum_{j=1}^K \mathbbm{1}\{x_{j,\ell} = x_{k,\ell}\} p_{jg}^t ~ \forall \ell,k,g$. 
Note that I know $C \left(\{\alpha_g^{n_t}\}, \{\beta_{k, \ell, g}^{n_t}\} \right)$ is nonempty, and is in fact a connected set. Because $\{p_{kg}\}$ satisfying $p_{kg} \geq 0$ define a nonempty closed set, I can choose the arbitrarily close solution $\{p_{kg}^t\}$ to be such that $p_{kg}^t \geq 0$ also holds.

From the equation $\alpha_g = \sum_{k=1}^K c_{kg}p_{kg}$ because $\{p_{kg}^t\}$ is arbitrarily close to $\{p_{kg}\}$, for the value of $n_t$ above there exists $\{c_{kg}^t\}$ arbitrarily close to $\{c_{kg}\}$ such that $\alpha_g^{n_t} = \sum_{k=1}^K c_{kg}^tp_{kg}^t$. Since there are infinitely many $\{c_{kg}^t\}$ that satisfy $\alpha_g^{n_t} = \sum_{k=1}^K c_{kg}^tp_{kg}^t$ for $K \geq 2$, I can choose $c_{kg}^t \in [y_\ell, y_u]$ for all $k,g$.

Finally if $\sum_{g=1}^G \P[G_i=g] p_{kg}^t \neq 0$ for a given $k$ then let $d_k^t = \frac{\sum_{g=1}^G \P[G_i=g] p_{kg}^tc_{kg}^t}{\sum_{g=1}^G \P[G_i=g] p_{kg}^t}$. If $\sum_{g=1}^G \P[G_i=g] p_{kg}^t = 0$ for a given $k$ then let $d_k^t = c_{k1}^t$. The restrictions of $\Gamma$ are satisfied for this choice of $\{d_k^t\}$. If for a given $k$, there exists $g$ with $p_{kg} \neq 0$ then for large enough $T$, $\sum_{g=1}^G \P[G_i=g] p_{kg}^t \neq 0$ for $t \geq T$. Thus by properties of limits $d_k^t \to d_k$ for that given $k$. If instead for a given $k$ all $p_{kg} = 0$, for all $t$ either $d_k^t = c_{k1}^t$ or $d_k^t$ is a convex combination of $c_{kg}^t$, but since $c_{kg}^t \to c_{kg} = d_k$ as $t \to \infty$ for each $g$, it follows that $d_k^t \to d_k$. Lower hemicontinuity, and therefore the result, follows.
\end{proof}

\subsection{Proof of Proposition \ref{prop:inference}} \label{app_proof_prop:inference}

Let $A_n = 1$ denote the event that all population analogs of the sample observations are jointly in their respective marginal confidence intervals. It is clear that for any sample size $n$, on $A_n$ the set of all $\{d_k\}$, $\{c_{kg}\}$, and $\{p_{kg}\}$ that satisfy the constraints of the optimization problems of Proposition \ref{prop:program_d} also satisfy the constraints of the optimization problems used to construct $\hat D_{CI,n}$. Thus if $A_n = 1$, $D \subseteq \hat D_{CI,n}$, that is, $\P[D \subseteq \hat D_{CI,n} \vert A_n = 1] = 1$ for all $n$.

By construction the population analog of every single sample observation is contained in its Bonferroni-corrected marginal confidence interval with joint probability that, in the limit as $n \to \infty$, is greater than $1-\alpha$. That is, $\lim_{n \to \infty} \P[A_n = 1] \geq 1-\alpha$. It follows that 
\begin{align*}
    \lim_{n \to \infty} \P[D \subseteq \hat D_{CI,n}] &\geq \lim_{n \to \infty} \P[D \subseteq \hat D_{CI,n} \vert A_n = 1] \P[A_n = 1] \geq 1-\alpha.
\end{align*}

\end{document}